\documentclass[prb,aps,10pt,showkeys,showpacs, twocolumn]{revtex4-2}
\usepackage{amsmath, amssymb, amsfonts, stmaryrd, wasysym, graphicx, multirow, color, textcomp, subfigure}
\usepackage{url, booktabs}
\usepackage[colorlinks=true, citecolor=blue, urlcolor=blue]{hyperref}
\usepackage[page, toc, titletoc, title]{appendix}
\usepackage[utf8]{inputenc}
\usepackage{comment}
\usepackage{soul, bm}
\usepackage[normalem]{ulem}
\usepackage[dvipsnames]{xcolor}
\setcitestyle{super, open={}, close={}}

\begin{document}

\newpage

\title{Emergence of Meron Kekulé lattices in twisted Néel antiferromagnets}

\author{Kyoung-Min Kim}
\email{kmkim@ibs.re.kr}
\affiliation{Center for Theoretical Physics of Complex Systems, Institute for Basic Science, Daejeon 34126, Republic of Korea}

\author{Se Kwon Kim}
\email{sekwonkim@kaist.ac.kr}
\affiliation{Department of Physics, Korea Advanced Institute of Science and Technology, Daejeon 34141, Republic of Korea}


\begin{abstract}
A Kekulé lattice is an exotic, distorted lattice structure exhibiting alternating bond lengths, distinguished from naturally formed atomic crystals. Despite its evident applicability, the formation of a Kekulé lattice from topological solitons in magnetic systems has remained elusive. Here, we propose twisted bilayer easy-plane Néel antiferromagnets as a promising platform for achieving a ``Meron Kekulé lattice"—a distorted topological soliton lattice comprised of antiferromagnetic merons as its lattice elements. We demonstrate that the cores of these merons are stabilized into the Kekulé-O pattern with different intracell and intercell bond lengths across moiré supercells, thereby forming a Meron Kekulé lattice. Moreover, the two bond lengths of the Meron Kekulé lattice can be fine-tuned by adjusting the twist angle and specifics of the interlayer exchange coupling, suggesting extensive control over the meron lattice configuration in contrast to conventional magnetic systems. These discoveries pave the way for exploring topological solitons with distinctive Kekulé attributes.
\end{abstract}

\maketitle

\begin{figure}[t!]
    \centering
    \includegraphics[width=0.48\textwidth]{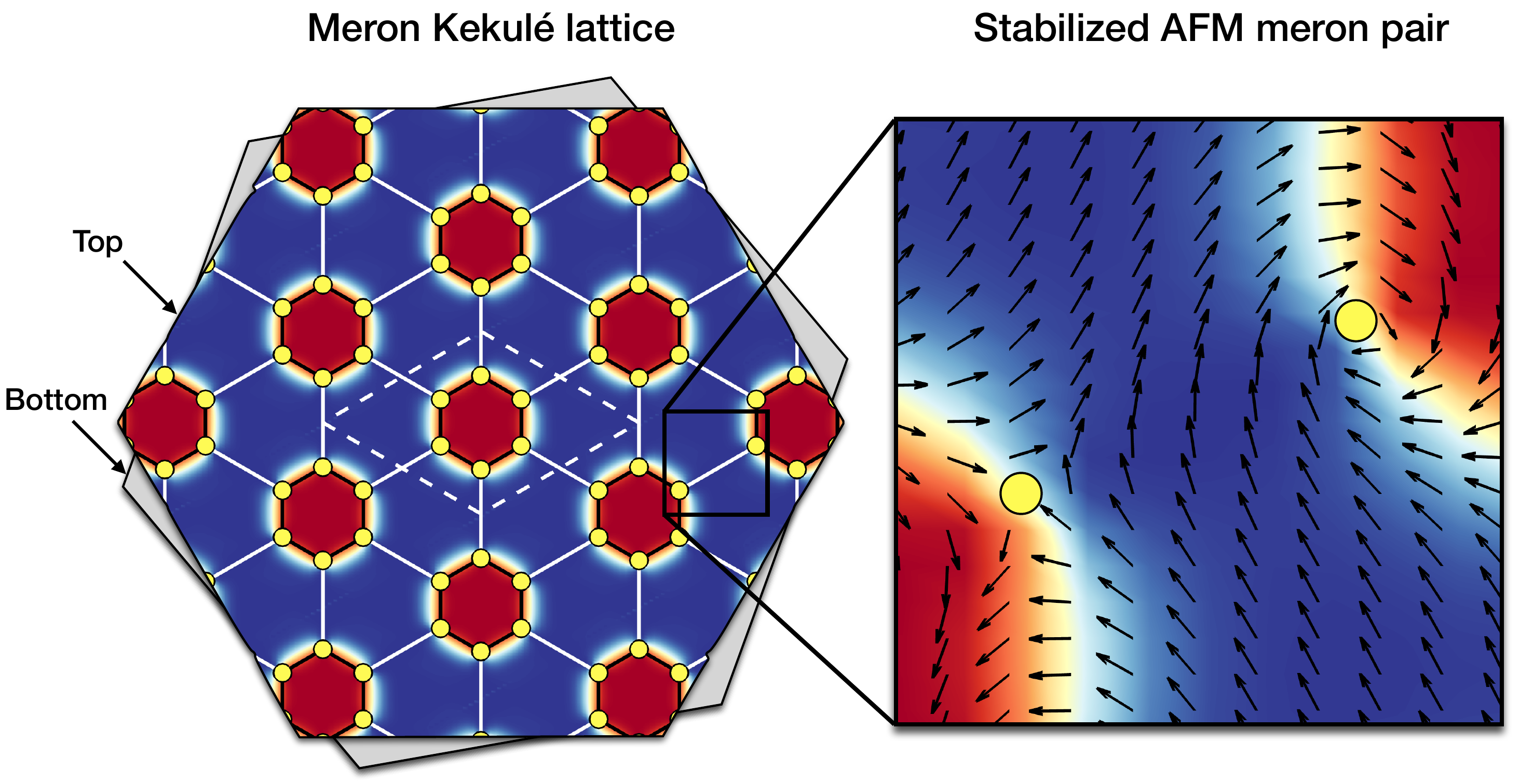}
    \caption{Schematic illustration of a Meron Kekulé lattice in a twisted bilayer Néel antiferromagnet. Yellow dots denote the cores of antiferromagnetic (AFM) merons that form a Kekulé lattice structure. Black solid lines denote intracell bonds between meron cores within the same moiré supercell, while white solid lines denote intercell bonds between meron cores across different supercells. The white dashed line depicts a single moiré supercell. Blue color indicates parallel alignment, while red indicates antiparallel alignment between Néel vectors across the top and bottom layers. In the magnified image, arrows represent the in-plane components of Néel vectors, depicting the winding texture of a meron pair.}
    \label{fig1}
\end{figure}

A Kekulé lattice is a distorted lattice structure distinguished by alternating bond lengths. This lattice distortion has been suggested to give rise to intriguing physical phenomena not observed in undistorted lattices \cite{PhysRevB.62.2806, PhysRevLett.98.186809, PhysRevB.80.233409, PhysRevB.80.205319, PhysRevB.81.085105, PhysRevB.83.201401, PhysRevB.88.245123, Mu2022, Gutierrez2016, PhysRevLett.119.255901, PhysRevLett.126.206804, PhysRevLett.131.266501, KAWARABAYASHI2021168440, PhysRevB.109.115131, 10.1063/1.5133091, PhysRevLett.124.236404, doi:10.7566/JPSJ.93.033703, Zhang_2023, Mohammadi_2021, PhysRevLett.119.255901, Gamayun_2018, Gao2020, Wei:22, Huang_2024, PhysRevApplied.11.044086, PhysRevB.81.085105, PhysRevLett.98.186809, PhysRevB.62.2806, Ma2018}. Notable examples include charge fractionalization \cite{PhysRevB.62.2806, PhysRevLett.98.186809}, chiral-symmetry breaking \cite{PhysRevB.80.233409, PhysRevB.81.085105, PhysRevB.83.201401, PhysRevB.88.245123, PhysRevB.80.205319, Mu2022, Gutierrez2016}, and emergence of flat bands \cite{PhysRevLett.131.266501} in graphene. Additionally, the appearance of topological band structures has been demonstrated in photonic \cite{Gao2020, Wei:22} and phononic crystals \cite{Huang_2024, PhysRevApplied.11.044086}. Experimentally, the Kekulé lattice has been realized in graphene \cite{Gutierrez2016, Ma2018, PhysRevLett.126.206804}. However, the concept of a Kekulé lattice can extend beyond these systems, holding promise for broader applications across various physical systems with distinct lattice elements. Particularly fascinating is the potential creation of a Kekulé lattice from topological solitons in magnetic systems, which would introduce a novel perspective, contrasting with the traditionally observed Bravais lattice forms of topological solitons, such as the triangular lattice seen in the Abrikosov vortex lattice in superconductors \cite{ESSMANN1967526} and the skyrmion lattice in magnetic systems \cite{doi:10.1126/science.1166767}. This intriguing possibility has remained unexplored, representing a pivotal piece yet to be uncovered within the realms of Kekulé physics.

In this study, we discover the creation of a Kekulé lattice from antiferromagnetic merons in twisted bilayer easy-plane Néel antiferromagnets. By conducting atomistic spin simulations on these magnets, we demonstrate that the cores of the antiferromagnetic merons are stabilized due to the moiré-induced spatial modulation of interlayer exchange coupling \cite{Kim2024}. Notably, the cores of these merons form a honeycomb-lattice-like structure across moiré supercells, exhibiting different intracell and intercell bond lengths (Figure~\ref{fig1}). This distorted structure deviates from commonly observed Bravais-lattice forms \cite{doi:10.1126/science.1166767, Yu2012, Yu2018, doi:10.1126/sciadv.abm7103, Peng2018}, yet aligning with the known characteristics of the Kekulé-O pattern \cite{Gutierrez2016}. Hence, we term our distorted meron lattice a ``Meron Kekulé lattice." We illustrate that the two bond lengths of the Meron Kekulé lattice can be fine-tuned by manipulating the twist angle and specifics of interlayer exchange coupling, offering a high degree of control over the stable meron lattice configuration. These discoveries present a Meron Kekulé lattice as a new classification within Kekulé lattices, introducing a novel pathway to attain this structure. Moreover, the stability and adjustability of this distorted structure present intriguing possibilities for exploring topological solitons with precise control—a capability rarely achievable within conventional magnetic systems \cite{doi:10.1126/science.1166767, Yu2012, Yu2018, doi:10.1126/sciadv.abm7103, Peng2018, doi:10.1126/science.289.5481.930, PhysRevLett.108.067205, PhysRevLett.110.177201, PhysRevB.94.014433, 10.1063/1.1483386, doi:10.1126/sciadv.aat3077, VanWaeyenberge2006, Ruotolo2009, PhysRevB.79.060407, Chmiel2018, Yu2018, Gao2019, Lu2020, Augustin2021}.

\begin{figure*}[t!]
    \centering
    \includegraphics[width=.97\textwidth]{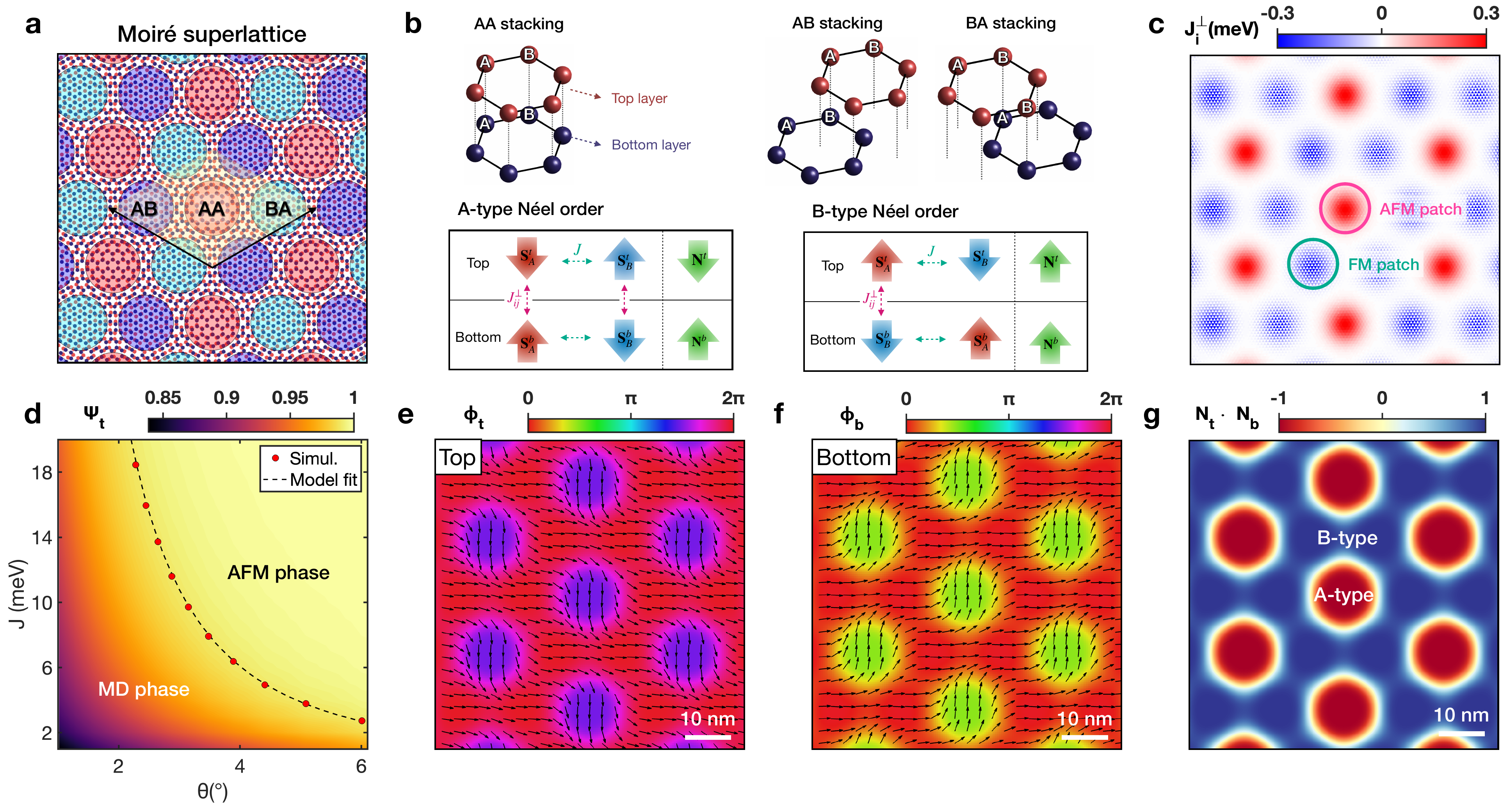}
    \caption[AFM domain array]{\textbf{Antiferromagnetic domain array.} \textbf{a} Moiré superlattice for a twist angle of $\theta=5.08$\textdegree{}. Colored circles highlight distinct local stacking patterns: AA (red), AB (blue), and BA (cyan). The yellow rhombus and black arrows denote the unit cell and lattice vectors of the moiré superlattice, respectively. \textbf{b} Schematic illustration depicting different local spin configurations (A-type and B-type Néel orders) preferred in each stacking order. Here, red and blue arrows represent spins on the A and B sublattice, respectively. Green arrows depict Neél vectors in each layer. \textbf{c} Interlayer exchange energy map ($J_{i}^{\perp}$) within the B-type order. Here, red and blue colors indicate the preference for the A-type and B-type orders, respectively. \textbf{d} Zero temperature magnetic phase diagram as a function of twist angle ($\theta$) and intralayer nearest-neighbor exchange interaction ($J$), displaying AFM and magnetic domain (MD) phases. The order parameter $\Psi_t$ is defined as $\Psi_t = \frac{1}{N_t}\left|\sum_i \bm{N}_{i}^{t}\right|$ ($\bm{N}_{i}^{t}$: a Néel vector on each $i$-site, $N_t$: the number of sites on the top layer), where $\Psi_t=1$ indicates the AFM phase, and $\Psi_t<1$ signifies the MD phase. The markers depict the boundary between the two phases. The dashed line represents a fitting curve defined as $J = 162/\theta^2$. \textbf{e}–\textbf{g} AFM domain array configuration of the MD phase. In \textbf{e}–\textbf{f}, the color scales denote the phase angles ($\phi_{t,b}$) of the normalized Néel vectors $\bm{N}_{t,b}=(\cos{\phi_{t,b}},\sin{\phi_{t,b}},0)$ in the top and the bottom layers, respectively. The arrows illustrate the direction of the Néel vectors in the plane. In \textbf{g}, the relative orientation of the Néel vectors between the two layers ($\bm{N}_t\cdot\bm{N}_b$), where red (blue) represents the A-type (B-type) Néel order. }
    \label{fig2}
\end{figure*}

We construct twisted bilayer Néel antiferromagnets by rotating two magnetic layers in a honeycomb lattice with a relative twist angle (Figure~\ref{fig2}\textbf{a}). These twisted magnets can be described using a Heisenberg spin model given by
\begin{equation} \label{eq:spinH}
\begin{aligned}
    H = & \; J\sum_{l=t,b} \sum_{\langle i, j\rangle} \bm{S}_{i}^{l} \cdot \bm{S}_{j}^{l} + A \sum_{l=t,b} \sum_{i} \big(\bm{S}_{i}^{l}\cdot\hat{\bm{z}}\big)^2  \\
    & + \sum_{i, j} J_{ij}^{\perp} \bm{S}_{i}^{t} \cdot \bm{S}_{j}^{b}.
\end{aligned}
\end{equation}
Here, $\bm{S}_{i}^{l}$ represents the spin at site $i$ on the top layer ($l=t$) and the bottom layer ($l=b$). The parameter $J=1.0$ meV represents the intralayer AFM exchange interactions between nearest-neighbor spins. $A=0.2$ meV represents the single-ion anisotropy energy favoring in-plane spin orientations. $J_{ij}^{\perp}$ represents the interlayer AFM exchange interactions that modulate as a function of the coordinate displacements $\bm{r}_{ij}^{tb}=\bm{r}_i^{t}-\bm{r}_j^{b}$ between two spins at $i$- and $j$- sites on the top and bottom layers, respectively. To describe the decaying behavior of $J_{ij}^{\perp}$ as a function of $|\bm{r}_{ij}^{tb}|$, we employ the following exponential function \cite{Tong2018, PhysRevB.103.L140406, Ghader2022, Kim2023, PhysRevB.108.L100401}: 
\begin{equation} \label{eq:J_perp}
    J_{ij}^{\perp} = J^{\perp}_0 \exp\Big[ - \alpha \Big(\big|\bm{r}_{ij}^{tb}\big|/d -1\Big)\Big].
\end{equation}
Here, the parameter $J_0^{\perp}$ represents the maximum value of $J_{ij}^{\perp}$ at $|\bm{r}_{ij}^{tb}|=d$, where $d$ denotes the perpendicular interlayer distance; the parameter $\alpha$ describes the decay rate of $J_{ij}^{\perp}$ with respect to the increase of $|\bm{r}_{ij}^{tb}|$. The twist angle \(\theta\) defines the size of the moiré superlattice \(L\) through the relationship: \(L \approx \frac{\sqrt{3}a}{\theta}\) (\(\theta\): in radians). We choose $\alpha=15$, $J^{\perp}_0 = 0.3~\textrm{meV}$, $d = 7~\textrm{\AA{}}$, and $a = 4~\textrm{\AA{}}$, aligned with typical values observed in diverse vdW magnetic materials \cite{Wang2022}. We use the twist angle $\theta=1.61$\textdegree{}. These parameter values are utilized throughout this study unless explicitly specified.

The moiré superlattice of the twisted bilayer encompasses various local stacking patterns, such as AA, AB, and BA, within its supercell (Figure~\ref{fig2}\textbf{a}). This stacking modulation leads to diverse spin alignments between the top and bottom layers \cite{Tong2018}. Two characteristic alignment patterns emerge:
\begin{equation}
\begin{aligned}
    \textrm{A-type Néel order: } \bm{N}_i^{t} \cdot \bm{N}_i^{b} = - 1, \\
    \textrm{B-type Néel order: } \bm{N}_i^{t} \cdot \bm{N}_i^{b} = + 1. \label{eq:Neel_orders}
\end{aligned}
\end{equation}
Here, $\bm{N}_i^{t}$ and $\bm{N}_i^{b}$ represent the Néel vectors on the top and bottom layers, respectively, defined as:
\begin{equation}
    \bm{N}_{i}^{t,b} = \frac{\bm{S}_{i,A}^{t,b}-\bm{S}_{i,B}^{t,b}}{\big|\bm{S}_{i,A}^{t,b}-\bm{S}_{i,B}^{t,b}\big|}. \label{eq:Neel_vectors}
\end{equation}
In the AA stacking regions, the A-type order predominates since spins on the same sublattice in the top and bottom layers (e.g., $\bm{S}_{i,A}^{t}$ and $\bm{S}_{i,A}^{b}$ for the sublattice A) are coupled through the AFM coupling $J_{ij}^{\perp}$ (the left panel of Figure~\ref{fig2}\textbf{b}). Conversely, the AB stacking regions favor the B-type order, as $\bm{S}_{i,A}^{t}$ and $\bm{S}_{i,B}^{b}$ are primarily coupled, while $\bm{S}_{i,B}^{t}$ and $\bm{S}_{i,A}^{b}$ remain effectively decoupled due to their large separations (the middle panel). The BA stacking regions prefer the B-type order with similar considerations (the right panel). The map of local interlayer exchange energy, computed as $J_{i}^{\perp} = \sum_{j} J_{ij}^{\perp} \bm{S}_i^{t}\cdot\bm{S}_j^{b}$, where $\bm{S}_{i,A}^{t}=\bm{S}_{i,A}^{b}=(1,0,0)$ and $\bm{S}_{i,B}^{t}=\bm{S}_{i,B}^{b}=(-1,0,0)$ reveals three distinct local regions (Figure~\ref{fig2}\textbf{c}): (i) ``AFM patches" preferring the A-type order, as indicated by their AFM coupling character between two Néel vectors (i.e., $J_{i}^{\perp} > 0$), (ii) ``FM patches" preferring the B-type order, as indicated by their FM coupling character (i.e., $J_{i}^{\perp} < 0$), and (iii) a neutral intermediate region lacking any specific preferred order, exhibiting negligible coupling (i.e., $J_{i}^{\perp} \approx 0$).

Our atomistic spin simulations on Eq.~\eqref{eq:spinH}, conducted using an iterative optimization method \cite{PhysRevB.108.L100401, Kim2023, Kim2024}, reveal a zero temperature magnetic phase diagram shown in Figure~\ref{fig2}\textbf{d}. For detailed methods, refer to Supporting Information (SI) \cite{SI}. This diagram displays two distinct magnetic phases: AFM and magnetic domain (MD) phases. The AFM phase displays a uniform spin configuration characterized by the B-type Néel order. In contrast, the MD phase exhibits a nonuniform configuration (Figure~\ref{fig2}\textbf{e}–\textbf{g}), where the A-type Néel order emerges within each AFM patch, forming circular-shaped domains, whereas the B-type Néel order persists outside these patches (Figure~\ref{fig2}\textbf{g}). We term this spin configuration an ``AFM domain array" \cite{Kim2024}. The AFM domain array configuration minimizes $J_i^\perp$ across both the AFM and FM patch regions. This energy reduction outweighs domain wall energy in a small twist angle regime \(\theta < \theta_c\), where $\theta_c = C\sqrt{\bar{J}_{\perp}/J}$ and $\bar{J}_\perp=0.3$ meV and $C=23.24$  \cite{Kim2023, PhysRevB.108.L100401, Kim2024}, resulting in the emergence of the MD phase as the ground state (Figure~\ref{fig2}\textbf{d}).

\begin{figure*}[t!]
    \centering
    \includegraphics[width=.9\textwidth]{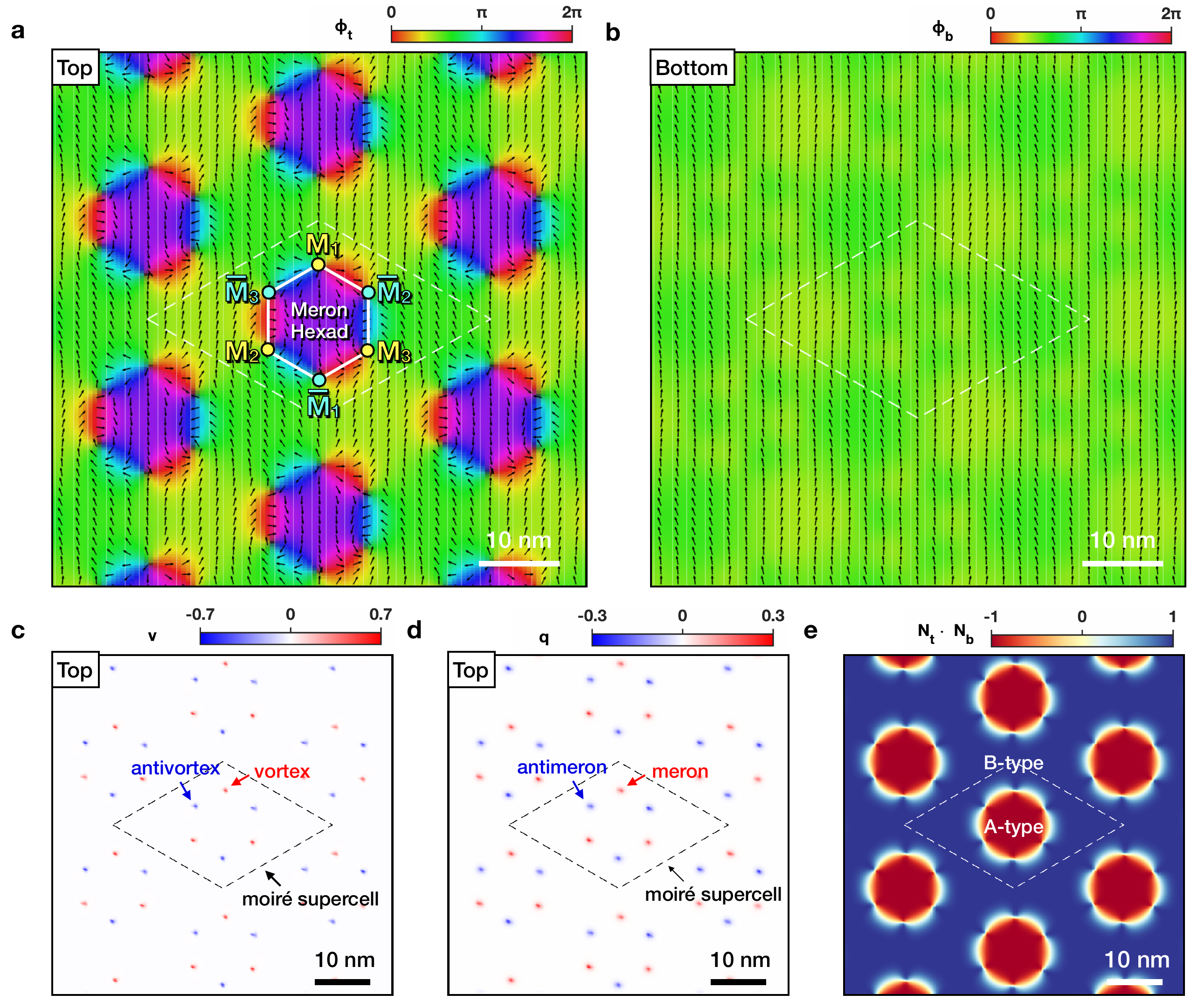}
    \caption{\textbf{Meron Kekulé lattice (MK) configuration.} \textbf{a} Néel vectors on the top layer ($\bm{N}_t$), showing three merons ($\textrm{M}_{1}$, $\textrm{M}_{2}$, $\textrm{M}_{3}$) and three antimerons ($\overline{\textrm{M}}_{1}$, $\overline{\textrm{M}}_{2}$, $\overline{\textrm{M}}_{3}$). \textbf{b} Néel vectors on the bottom layer ($\bm{N}_b$), displaying almost homogeneous alignment. In \textbf{a}–\textbf{b}, the color scales denote the phase angles ($\phi_{t,b}$) of the normalized Néel vectors $\bm{N}_{t,b}=(\sin{\theta_{t,b}}\cos{\phi_{t,b}},\sin{\theta_{t,b}}\sin{\phi_{t,b}},\cos{\theta_{t,b}})$ in the top and the bottom layers, respectively. The black color denotes the out-of-plane component in the direction perpendicular to the plane (i.e., $\bm{N}_t=(0,0,1)$). The arrows depict the interpolation of the in-plane components of the Néel vectors. \textbf{c} Vortex density ($v$) map corresponding to \textbf{a}, with red (blue) color indicating a positive (negative) winding number. \textbf{d} Skyrmion density ($q$) map corresponding to \textbf{a}, with red (blue) color indicating a positive (negative) skyrmion number. \textbf{e} The relative orientation of the Néel vectors between the two layers ($\bm{N}_t\cdot\bm{N}_b$), with blue (red) indicating parallel (antiparallel) alignment. In each panel, the dashed line depicts a supercell. }
    \label{fig3}
\end{figure*}

We demonstrate that the AFM domain array, depicted in Figure~\ref{fig2}\textbf{e}–\textbf{g}, can accommodate merons as topological defects around its domain boundary, leading to a Meron Kekulé lattice. Merons are vortexlike topological solitons carrying half-skyrmion numbers \cite{Yu2018, Gao2019}, typically found in easy-plane magnets \cite{doi:10.1126/science.289.5481.930, PhysRevLett.108.067205, PhysRevLett.110.177201, PhysRevB.94.014433, 10.1063/1.1483386, doi:10.1126/sciadv.aat3077, VanWaeyenberge2006, Ruotolo2009, PhysRevB.79.060407, Chmiel2018, Yu2018, Gao2019, Lu2020, Augustin2021}. The local density profile for the skyrmion number is calculated as:
\begin{equation}
    q(x,y)=\frac{1}{4\pi}\bm{N}(x,y)\cdot(\partial_{x}\bm{N}(x,y)\times \partial_{y}\bm{N}(x,y)), \label{eq:skyr_dens}
\end{equation}
where $\bm{N}(x,y)$ represents the normalized interpolation of the Néel vectors. Furthermore, in-plane swirling textures away from the cores of merons are characterized by vorticity. The local density profile for the vorticity is calculated as \cite{PhysRevB.99.180402}:
\begin{equation}
    v(x,y)=\frac{1}{\pi}\hat{\bm{z}}\cdot(\partial_{x}\bm{N}(x,y)\times\partial_{y}\bm{N}(x,y)). \label{eq:vort_dens}
\end{equation}

Figure~\ref{fig3} presents a stable spin configuration obtained through the relaxation of a random initial configuration. On the top layer, the Néel vectors exhibit six-fold in-plane swirling textures within a supercell, which are labeled by $\textrm{M}_{1}$, $\textrm{M}_{2}$, $\textrm{M}_{3}$, $\overline{\textrm{M}}_{1}$, $\overline{\textrm{M}}_{2}$, and $\overline{\textrm{M}}_{3}$ (Figure~\ref{fig3}\textbf{a}). The map for $v(x,y)$, presented in Figure~\ref{fig3}\textbf{c}, displays positive and negative values for ($\textrm{M}_{1}$, $\textrm{M}_{2}$, $\textrm{M}_{3}$) and ($\overline{\textrm{M}}_{1}$, $\overline{\textrm{M}}_{2}$, $\overline{\textrm{M}}_{3}$), respectively, identifying the former as vortices and the latter as antivortices. Furthermore, the map for $q(x,y)$, depicted in Figure~\ref{fig3}\textbf{d}, demonstrates nonvanishing skyrmion densities with sign alternation for each core. These densities are locally integrated around each core region: $Q=\int dxdyq(x,y)$, yielding skyrmion numbers $Q\approx-1/2$ for the vortices and $Q \approx +1/2$ for the antivortices (for detailed methods, refer to SI \cite{SI}). These half-skyrmion number characteristics identify ($\textrm{M}_{1}$, $\textrm{M}_{2}$, $\textrm{M}_{3}$) as merons and ($\overline{\textrm{M}}_{1}$, $\overline{\textrm{M}}_{2}$, $\overline{\textrm{M}}_{3}$) as antimerons \cite{Yu2018, Gao2019}.

\begin{figure*}
    \centering
    \includegraphics[width=.97\textwidth]{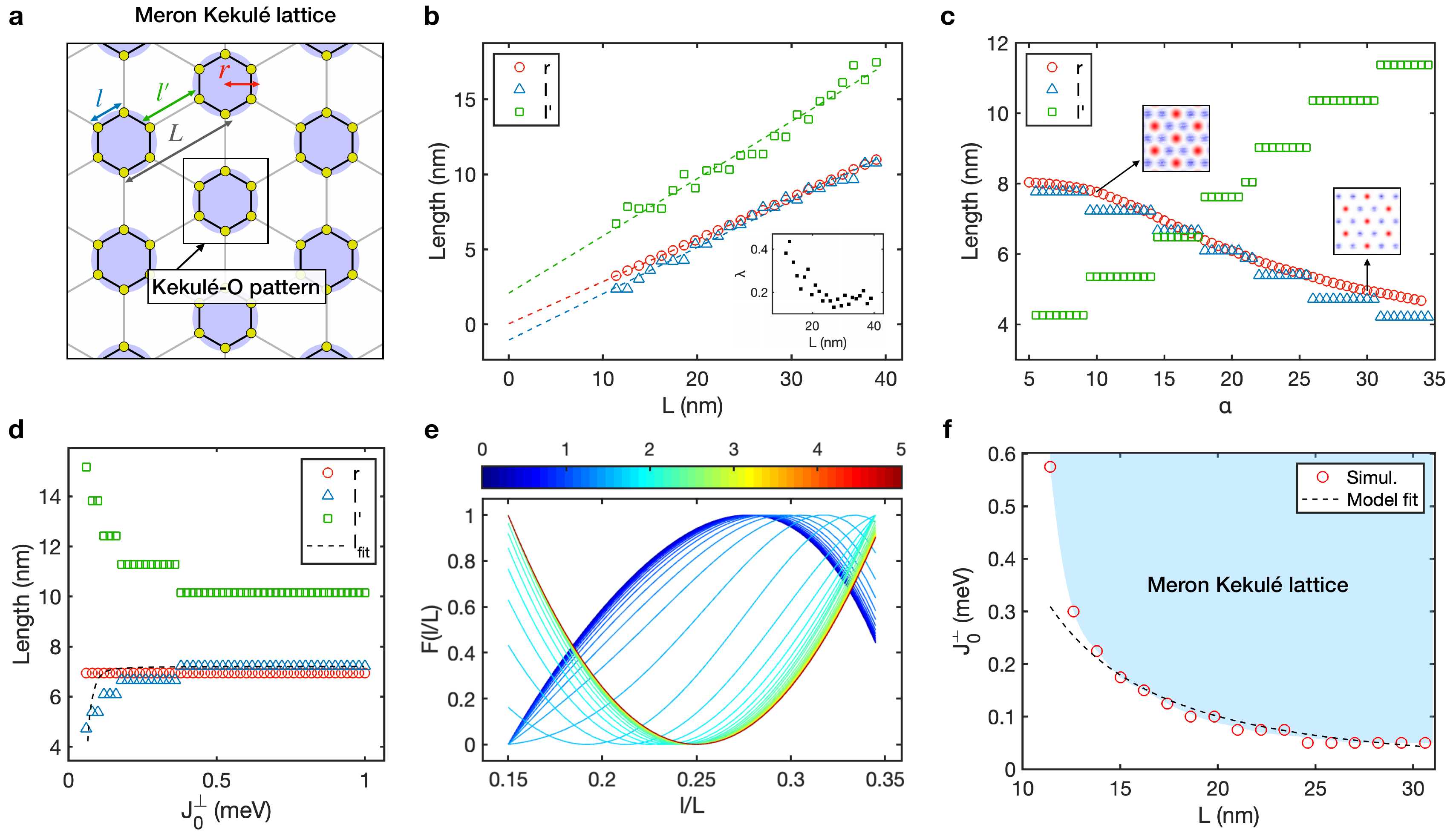}
    \caption{\textbf{Manipulation of the Kekulé structure through parameter adjustments.} \textbf{a} Schematic illustration of the Meron Kekulé lattice structure. Yellow dots denote the position of the meron cores. Black lines denote intracell bonds between meron cores within a Meron hexad, while light gray lines denote intercell bonds between meron cores across different Meron hexads. $l$ and $l'=l-2L$ denote the intracell and intercell bond lengths, respectively. $r$ denotes the size of AFM patches marked by puple circles. \textbf{b}–\textbf{d} Manipulation of the Kekulé lattice structure through three different parameters: \textbf{b} the moiré superlattice size ($L$), \textbf{c} the decay rate parameter of interlayer exchange interactions $J_{ij}^\perp$ ($\alpha$), and \textbf{d} the maximum strength of $J_{ij}^\perp$ ($J_0^\perp$). In \textbf{b}, the inset shows the value of the Kekulé constant ($\lambda=\frac{l'-l}{L}$). The dashed lines represent empirical fitting curves, described by \(r=0.28L+0.05\), \(l = 0.31L - 1.04\), and \(l' = 0.38L + 2.09\). In \textbf{c}, the insets display $J_i^\perp$ for $\alpha=10$ (left) and $\alpha=30$ (right), respectively. The same color scale used in Figure~\ref{fig2}\textbf{c} is applied in the insets. In \textbf{d}, the dashed line represents the fitting curve for the $l$ values. \textbf{e} Effective energy function $E(l)$ for the MK state, plotted in a normalized form: $F(l/L)=\frac{E(l)-\textrm{min}[E(l)]}{\textrm{max}[E(l)]-\textrm{min}[E(l)]}$. Here, `min' and `max' indicate the minimum and maximum values, respectively. Different colors indicate distinct values of $\log_{10}(p)$, where $p$ represents the dimensionless ratio between the Coulomb interactions and the moiré field, which represent the two effective interactions influencing merons in the effective model. The parameter $r/L=0.25$ is used. \textbf{f} Illustration of the stabilization condition of the MK state as a function of $L$ and $J_0^\perp$. The MK state is stable in the blue region while unstable in the devoid region. The markers represent simulation data points, while the dashed line is a fitting curve defined as $J_0^\perp=40/L^2$.}
    \label{fig4}
\end{figure*}

The cores of the merons and antimerons form six vertices of a hexagon along the boundary of the AFM patch (Figure~\ref{fig3}\textbf{a}). We refer to this distinctive configuration as a ``Meron hexad." Despite its intricate texture, the Meron hexad attains an AFM domain array configuration that minimizes interlayer exchange energy across the bulk region (Figure~\ref{fig3}\textbf{e}). Furthermore, the constituent meron cores of Meron hexads form a honeycomb-lattice-like pattern, where the distance between two nearest-neighbor meron cores from the same hexad is shorter than that between cores from adjacent Meron hexads (Figure~\ref{fig3}\textbf{a}). This distorted structure is referred to as a Kekulé lattice with a Kekulé-O pattern \cite{Gutierrez2016}. Thus, we term this distorted meron lattice a ``Meron Kekulé lattice (MK)." The Kekulé structure of the MK state is characterized by two key length parameters: the intracell bond length within the same Meron hexad ($l$) and the intercell bond length across different Meron hexads ($l'=L-2l$). The distinction between $l$ and $l'$ indicates the deviation from the conventional honeycomb lattice, giving rise to an alternative Kekulé structure. To quantify this deviation, we introduce a dimensionless constant $\lambda$, referred to as the ``Kekulé constant:"
\begin{equation}
    \lambda=\frac{l'-l}{L}=1-\frac{3l}{L}, \label{eq:kekule_const}
\end{equation}
where \(\lambda = 0\) and \(\lambda > 0\) indicate regular honeycomb and Kekulé lattices, respectively.

We illustrate how variations in the parameters of the spin model influence the values of \(l\) and \(l'\), as well as the AFM patch size (\(r\)). Figure~\ref{fig4}\textbf{b} shows the evolution of \(l\) and \(l'\) as functions of \(L\). We first observe a linear increase in the AFM patch size (\(r\)) with \(L\), which is evident since the increase in \(L\) signifies expansions in all local stacking regions, including AFM patches. The increase in \(r\) correlates with an increase in \(l\), as the minimization of interlayer exchange energy dictates \(l = r\) \cite{Kim2024}. In a large \(L\) regime, this relationship holds precisely; however, it progressively deviates in a small \(L\) regime. This deviation from \(l = r\) contributes to the enhancement of Kekulé distortion, represented by \(\lambda\) (see the inset). The linear increase in \(l'\) directly follows the increases in \(L\) and \(l\), as described by the relationship \(l' = L - 2l\).

We find that an increase in \(\alpha\) leads to the contraction of \(r\), resulting in a decrease in \(l\) while adhering to the relationship \(l \approx r\). Meanwhile, \(l'\) increases according to \(l' = L - 2l\) with a fixed \(L\) (Figure~\ref{fig4}\textbf{c}). The contraction of \(r\) is attributed to significant suppression of \(J_{ij}^{\perp}\) for large values of \(|\bm{r}_{ij}^{tb}|\) away from the center of an AFM patch (see the insets). Moreover, an increase in \(J_0^\perp/J\) initially leads to an increase in \(l\) when \(J_0^\perp/J\) is sufficiently small (Figure~\ref{fig4}\textbf{d}). However, \(l\) reaches its maximum value \(l_{\text{max}} \approx r\), and further changes in \(J_0^\perp/J\) do not affect \(l\). We also find that reducing the ratio \(d/a\) lowers \(l\) by decreasing \(r\), similar to the effects of \(\alpha\). Furthermore, varying \(A/J\) does not alter \(l\), while extreme values lead to a transition to other magnetic states. Further details regarding this analysis of \(d/a\) and \(A/J\) can be found in SI \cite{SI}.

\begin{table}[t!]
    \centering
    \begin{tabular}{cccc|ccccc} \toprule   
    $\theta$ (\textdegree{}) & $\alpha$ & $d/a$ & $J_0^\perp$ & $l$ (nm) & $l'$ (nm) & $L$ (nm) & $\lambda$ \\ \midrule
    1.61 & 15 & 1.75 & 0.3 & 6.66 & 11.28 & 24.60 & 0.19 \\ \midrule
    1.05 & 15 & 1.75 & 0.3 & 10.74 & 16.32 & 37.80 & 0.15 \\ 
    2.28 & 15 & 1.75 & 0.3 & 4.23 & 8.94 & 17.40 & 0.27 \\ \midrule
    1.61 & 8  & 1.75 & 0.3 & 7.77 & 9.06  & 24.60 & 0.05 \\ 
    1.61 & 25 & 1.75 & 0.3 & 5.39 & 13.82 & 24.60 & 0.34 \\ \midrule
    1.61 & 15 & 1.40 & 0.3 & 6.09 & 12.42 & 24.60 & 0.26 \\ 
    1.61 & 15 & 2.00 & 0.3 & 7.22 & 10.16 & 24.60 & 0.12 \\ \midrule
    1.61 & 15 & 1.75 & 0.1 & 5.38 & 13.83 & 24.60 & 0.34 \\
    1.61 & 15 & 1.75 & 0.5 & 7.22 & 10.15 & 24.60 &  0.12 \\
    \bottomrule
    \end{tabular}
    \caption{Examples of Kekulé structures (characterized by four parameters $l, l', L$, and $\lambda$) in the MK state with various parameters. The first row corresponds to the MK state shown in Figure~\ref{fig3}. Subsequent rows illustrate the changes in the Kekulé structures induced by the variations in the twist angle ($\theta$), the decay rate of interlayer exchange coupling ($\alpha$), the lattice length ratio ($d/a$), and the interlayer exchange coupling strength ($J_0^\perp$) from the first-row case. Here, $a$ and $d$ denote the nearest-neighbor distance within the same layer and the perpendicular interlayer distance, respectively. $a$ is set to 4 Å.}
    \label{tab1}
\end{table}

Table~\ref{tab1} presents Meron Kekulé lattices at various length scales (\(l\), \(l'\), and \(L\)). These structures can be created by manipulating the twist angle (\(\theta\)) \cite{Song2021, Xu2022, Xie2023} or by modifying the parameters of interlayer exchange coupling (\(\alpha\), \(d/a\), and \(J_0^\perp/J\)). The latter approach can be implemented using different antiferromagnetic materials or external controls such as gate voltage, intercalation, adsorption, strain, or surface modification \cite{Jiao_2023}. The resulting Kekulé lattices exhibit a wide range of distortion, with \(\lambda\) values ranging from 0.05 to 0.34, highlighting the adaptability of the Meron Kekulé lattice.

We attribute the intricate behavior of $l$ observed in Figure~\ref{fig4}\textbf{b}–\textbf{d} to the interplay between two magnetic interactions influencing merons: (i) attractive Coulomb interactions between merons and antimerons with opposite vorticities \cite{Hubert1998}, induced by the intralayer exchange interactions, and (ii) an effective moiré field generated by the bottom layer through interlayer exchange interactions \cite{Kim2024}. To elucidate their interplay, we utilize an energy function derived from an effective continuum model:
\begin{equation}
\begin{aligned}
E(l) = & \; C_1J\ln\Bigg[\sqrt{\frac{2l}{3L}}\frac{\big(L^6-l^6\big)^{2}\big(L^6-64l^6\big)}{L^6\big(L^6+27l^6\big)^{2}} \Bigg]  \\ 
	& + \frac{C_2J_0^{\perp}}{a^2} (l - r)^2. \label{eq:eff_pot} 
\end{aligned}
\end{equation}
Here, the first term describes the Coulomb interactions ranging up to the nearest-neighbor supercells, and the second term represents the moiré field. The constants $C_1$ and $C_2$ are numerical factors on the order of unity. The relative strength of these two interactions is represented by the dimensionless ratio: 
\begin{equation}
	p = \frac{2C_2J_0^{\perp}L^2}{C_1Ja^2}.
\end{equation}
The value of \(l\), where the MK state is stabilized, denoted as \(l_*\), can be determined by minimizing \(E(l)\) with respect to \(l\). When \(p\) is large, \(l_* \approx r\) (e.g., \(l \approx 0.25L\) for \(r = 0.25L\) in Figure~\ref{fig4}\textbf{e}), indicating that the moiré field term dominates. Conversely, a reduction in \(p\) enhances the role of the Coulomb interaction term, leading to a decrease in \(l_*\), as shown by the shifts in the local minimum of \(E(l)\) in Figure~\ref{fig4}\textbf{e}. We assert that this competition plays a crucial role in the intricate behavior of \(l\) observed in Figures~\ref{fig4}\textbf{b}–\textbf{d}. In particular, our analysis demonstrates that the \(l_*\) value derived from \(E(l)\) captures the qualitative features of the observed \(l\) values in the simulation, supporting this assertion (the dashed lines in Figure~\ref{fig4}\textbf{d}). For the derivation of \(E(l)\) and the fitting method for \(l\), refer to SI \cite{SI}.

The effective potential \(E(l)\) offers insights into the underlying mechanisms that support the stability of merons within the MK state. In conventional magnets, merons tend to be unstable due to attractive Coulomb interactions, which lead to pair annihilation during energy relaxation processes. However, in our twisted magnets, the moiré field mitigates this instability by imposing an energy penalty for the deviation of the meron cores from the AFM domain boundary, leading to the energetic stability of the MK state, shown by the convexity of \(E(l)\) (Figure~\ref{fig4}\textbf{e}). Furthermore, the effective model predicts that the stabilization of the MK state can be achieved by increasing either the interlayer exchange coupling (\(J_0^{\perp}\)) or the superlattice size (\(L\)), which is confirmed in our numerical simulations (Figure~\ref{fig4}\textbf{f}).

\begin{figure} [t!]
    \centering
    \includegraphics[width=.48\textwidth]{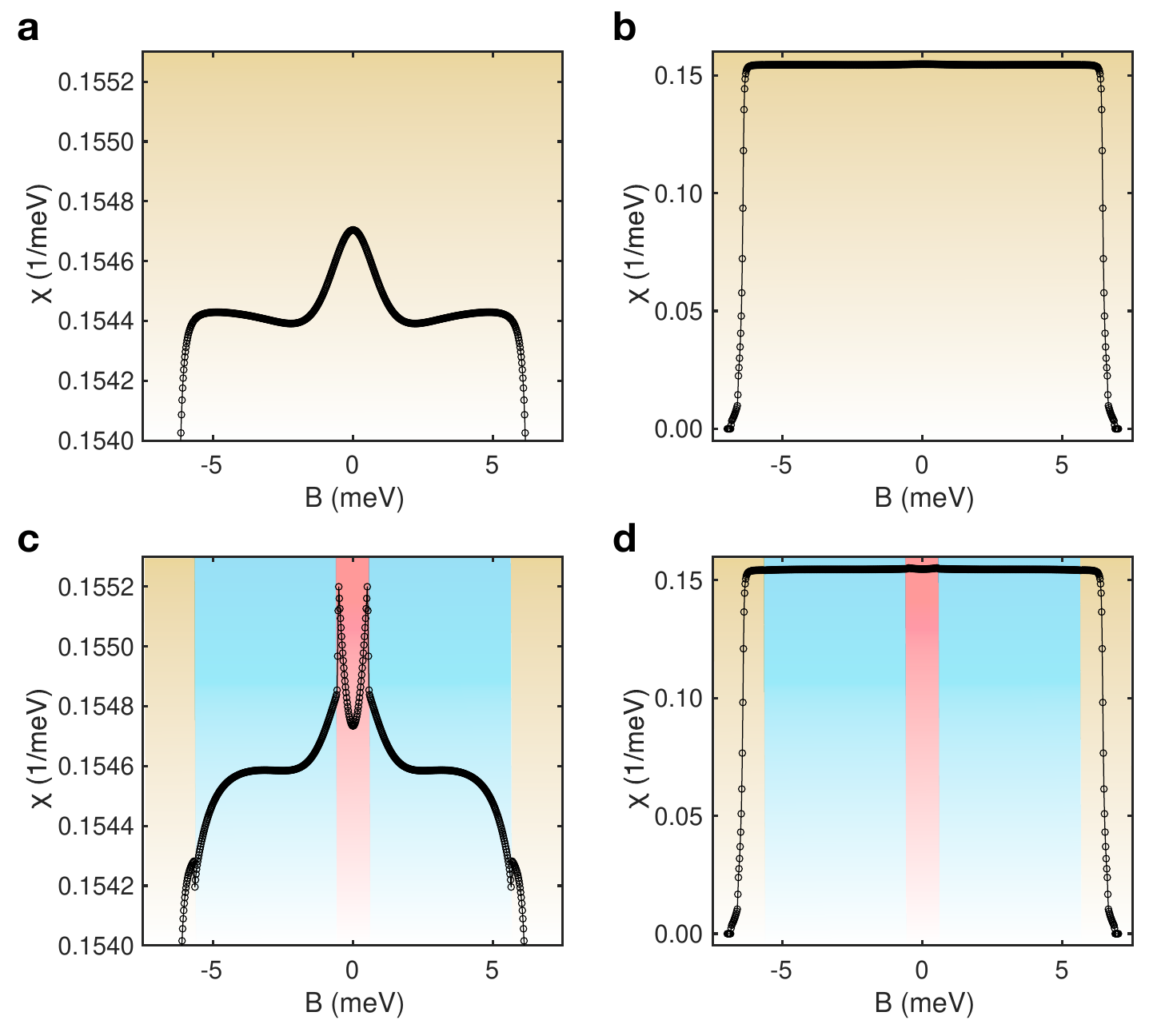}
    \caption{\textbf{a} Magnetic susceptibility ($\chi=dM/dB$) of the MD state with respect to the out-of-plane magnetization ($M$) as a function of an external magnetic field in the out-of-plane direction ($B$). \textbf{b} Zoom-out view of \textbf{a}. In \textbf{a}–\textbf{b}, the yellow color highlights the persistence of the MD state. \textbf{c} Magnetic susceptibility ($\chi$) of the MK state. Different colors indicate three different magnetic states: MK state (red), vortex state (blue), and MD state (yellow). \textbf{d} Zoom-out view of \textbf{c}. }
    \label{fig5}
\end{figure}

To investigate a potential means to distinguish the MD and MK states by bulk measurements, we analyze the magnetic response of the MD and MK states to external magnetic fields, incorporating the Zeeman term defined as:
\begin{equation}
    H_\textrm{Zeeman} = - \sum_{l=t,b} \sum_{i} \bm{B} \cdot \bm{S}_{i}^{l}, \label{eq:Zeeman}
\end{equation}
where $\bm{B} = B \hat{\bm{z}}$ denotes external magnetic fields in the out-of-plane magnetization. The initial MD and MK states, presented in Figures~\ref{fig3} and \ref{fig4}, are relaxed under external magnetic fields with continuously varying strength $B$. From the relaxed spin configurations for each state, the magnetic susceptibility, $\chi \equiv dM/dB$, is calculated as a function of $B$, where $M = \frac{1}{N} \sum_{i,l} \bm{S}_{i}^{l} \cdot \hat{\bm{z}}$ is the average magnetization in the out-of-plane magnetization. Figure~\ref{fig5}\textbf{a}–\textbf{b} presents $\chi$ of the MD state. Near $B = 0$, $\chi$ initially decreases as $|B|$ increases, while it shows an upturn in $|B|>2.22$ meV, exhibiting shoulder-like features around $|B|=4.88$ meV (Figure~\ref{fig5}\textbf{a}). Beyond these shoulders, $\chi$ sharply drops, and subsequently vanishes at $|B| > B_\textrm{sat.}=6.86$ meV (Figure~\ref{fig5}\textbf{b}), where $M$ saturates to $|M| = 1$.

Figure~\ref{fig5}\textbf{c}–\textbf{d} presents $\chi$ of the MK state. In the low-field regime ($|B| < B_{c1}=0.52$ meV), $\chi$ exhibits an increasing tendency with $|B|$ in contrast to the MD state. This behavior is attributed to the canted spin configurations in the core region that enhance the magnetization induced by the field. However, entering into the intermediate-field regime ($B_{c1} \leq |B| < B_{c2}= 5.62$ meV), $\chi$ undergoes a sharp drop and subsequently exhibits a decreasing tendency. Our findings reveal that at the threshold $|B| = B_{c1}$, the spin flipping occurs in the perpendicular direction for spins with out-of-plane components opposite to the field direction. The resulting state maintains in-plane swirling textures while lacking out-of-plane components in their Néel vectors. Consequently, this state retains only vorticities but loses half-skyrmion numbers characteristics, thus being named a ``vortex state" (for the detailed analysis of this state, refer to SI \cite{SI}). Finally, upon entering the high-field regime ($|B| \geq B_{c2}$), the vortex state transitions into the MD state through the annihilation of meron-antimeron pairs. At this regime, $\chi$ aligns with that of the MD state. We highlight the two anomalous discontinuities at the transition points ($|B|=B_{c1}$ and $|B|=B_{c2}$) as discernible signatures of the formation of merons in the MK state.

Aside from the configuration in Figure~\ref{fig3}, the MK state can manifest in a diverse array of \(2^{6}\) potential configurations, each defined by two possible polarities (\(p = \pm 1\)) for the six meron cores within a Meron hexad. This configurational diversity suggests that the MK state may be more favorable at finite temperatures compared to the MD state. We find that the expected occupation number of Meron hexads in each supercell, denoted as \(f_{\text{MH}}\), is \(f_{\text{MH}} = 0.32\) at \(\theta = 1.02^\circ\) and \(T = 40\) K, while \(f_{\text{MH}} = 0.49\) at \(\theta = 0.50^\circ\) and \(T = 20\) K. These results imply that achieving the MK state can be facilitated by either increasing the temperature or adjusting the twist angle to a lower value \cite{Jena2020}. Further details can be found in SI \cite{SI}.

We propose that our theory can be applied to CoPS\textsubscript{3}, which exhibits two essential factors for achieving the Meron Kekulé lattice: AFM intralayer exchange interactions and easy-plane magnetic anisotropy. Another potential application lies in XPS\textsubscript{3} (X = Mn, Ni, Fe) and MnPSe\textsubscript{3}, where their intrinsic easy-axis magnetic anisotropy can be potentially altered into easy-plane anisotropy through experimental control, such as strain, gate voltage, or surface adsorption \cite{Kim2020, PhysRevB.98.144411, XU2020126754, D0TC04049E, Tang2023}.

For experimental observations, we propose utilizing scanning magnetometry techniques with nitrogen-vacancy centers, as well as Lorentz transmission electron microscopy and magnetic transmission soft X-ray microscopy, to directly observe meron pairs ranging in size from 40 to 10 nm at different twist angles ($\theta =$ 0.5\textdegree{} to 2\textdegree{}). Indirect measurements can involve detecting the anomalous multiple superlattice periodicity manifested by different bond lengths meron through neutron scattering experiments. Additionally, anomalous kinks in the magnetization curve, as observed in Figure~\ref{fig5}, can offer valuable insights for such indirect measurements, serving as an indication of the presence of merons.

An intriguing research question to explore is how the formation of a Meron Kekulé lattice affects magnon band structures through its distorted lattice structure in vdW antiferromagnets. One potential outcome suggested by prior investigations is the creation of topological band structures by opening a band gap through hopping amplitude modulation \cite{Wei:22, Huang_2024, PhysRevApplied.11.044086}. Furthermore, exploring phonon-like collective-excitation modes stemming from meron core vibrations within the Meron Kekulé lattice presents a compelling avenue for investigating topological band structures, similar to the approach used for lattice-forming skyrmion cores \cite{PhysRevLett.119.077204, PhysRevB.98.180407}.

\section*{Acknowledgement}
K.-M.K. would like to thank Chang-Hwan Yi and Beom Hyun Kim for sharing their insights. K.-M.K. was supported by the Institute for Basic Science in the Republic of Korea through the project IBS-R024-D1. S.K.K. was supported by Brain Pool Plus Program through the National Research Foundation of Korea funded by the Ministry of Science and ICT (2020H1D3A2A03099291) and National Research Foundation of Korea funded by the Korea Government via the SRC Center for Quantum Coherence in Condensed Matter (RS- 2023-00207732). \\

\noindent \textbf{Supporting Information:} Spin configuration relaxation methods, derivation of the effective energy function of the MK state, alternation of intracell bond length with variation of $d/a$ and $A$, detailed analysis of the vortex state, classification of distinct MK states, estimation of occupation number of Meron Hexads at finite temperatures.

\noindent \textbf{Competing interests:} The authors declare that they have no competing interests.

\bibliography{ref.bib}

\begin{thebibliography}{66}%
\makeatletter
\providecommand \@ifxundefined [1]{%
 \@ifx{#1\undefined}
}%
\providecommand \@ifnum [1]{%
 \ifnum #1\expandafter \@firstoftwo
 \else \expandafter \@secondoftwo
 \fi
}%
\providecommand \@ifx [1]{%
 \ifx #1\expandafter \@firstoftwo
 \else \expandafter \@secondoftwo
 \fi
}%
\providecommand \natexlab [1]{#1}%
\providecommand \enquote  [1]{``#1''}%
\providecommand \bibnamefont  [1]{#1}%
\providecommand \bibfnamefont [1]{#1}%
\providecommand \citenamefont [1]{#1}%
\providecommand \href@noop [0]{\@secondoftwo}%
\providecommand \href [0]{\begingroup \@sanitize@url \@href}%
\providecommand \@href[1]{\@@startlink{#1}\@@href}%
\providecommand \@@href[1]{\endgroup#1\@@endlink}%
\providecommand \@sanitize@url [0]{\catcode `\\12\catcode `\$12\catcode
  `\&12\catcode `\#12\catcode `\^12\catcode `\_12\catcode `\%12\relax}%
\providecommand \@@startlink[1]{}%
\providecommand \@@endlink[0]{}%
\providecommand \url  [0]{\begingroup\@sanitize@url \@url }%
\providecommand \@url [1]{\endgroup\@href {#1}{\urlprefix }}%
\providecommand \urlprefix  [0]{URL }%
\providecommand \Eprint [0]{\href }%
\providecommand \doibase [0]{https://doi.org/}%
\providecommand \selectlanguage [0]{\@gobble}%
\providecommand \bibinfo  [0]{\@secondoftwo}%
\providecommand \bibfield  [0]{\@secondoftwo}%
\providecommand \translation [1]{[#1]}%
\providecommand \BibitemOpen [0]{}%
\providecommand \bibitemStop [0]{}%
\providecommand \bibitemNoStop [0]{.\EOS\space}%
\providecommand \EOS [0]{\spacefactor3000\relax}%
\providecommand \BibitemShut  [1]{\csname bibitem#1\endcsname}%
\let\auto@bib@innerbib\@empty
\bibitem [{\citenamefont {Chamon}(2000)}]{PhysRevB.62.2806}%
  \BibitemOpen
  \bibfield  {author} {\bibinfo {author} {\bibfnamefont {C.}~\bibnamefont
  {Chamon}},\ }\bibfield  {title} {\bibinfo {title} {Solitons in carbon
  nanotubes},\ }\href {https://doi.org/10.1103/PhysRevB.62.2806} {\bibfield
  {journal} {\bibinfo  {journal} {Phys. Rev. B}\ }\textbf {\bibinfo {volume}
  {62}},\ \bibinfo {pages} {2806} (\bibinfo {year} {2000})}\BibitemShut
  {NoStop}%
\bibitem [{\citenamefont {Hou}\ \emph {et~al.}(2007)\citenamefont {Hou},
  \citenamefont {Chamon},\ and\ \citenamefont {Mudry}}]{PhysRevLett.98.186809}%
  \BibitemOpen
  \bibfield  {author} {\bibinfo {author} {\bibfnamefont {C.-Y.}\ \bibnamefont
  {Hou}}, \bibinfo {author} {\bibfnamefont {C.}~\bibnamefont {Chamon}},\ and\
  \bibinfo {author} {\bibfnamefont {C.}~\bibnamefont {Mudry}},\ }\bibfield
  {title} {\bibinfo {title} {Electron fractionalization in two-dimensional
  graphenelike structures},\ }\href
  {https://doi.org/10.1103/PhysRevLett.98.186809} {\bibfield  {journal}
  {\bibinfo  {journal} {Phys. Rev. Lett.}\ }\textbf {\bibinfo {volume} {98}},\
  \bibinfo {pages} {186809} (\bibinfo {year} {2007})}\BibitemShut {NoStop}%
\bibitem [{\citenamefont {Cheianov}\ \emph {et~al.}(2009)\citenamefont
  {Cheianov}, \citenamefont {Sylju\aa{}sen}, \citenamefont {Altshuler},\ and\
  \citenamefont {Fal'ko}}]{PhysRevB.80.233409}%
  \BibitemOpen
  \bibfield  {author} {\bibinfo {author} {\bibfnamefont {V.~V.}\ \bibnamefont
  {Cheianov}}, \bibinfo {author} {\bibfnamefont {O.}~\bibnamefont
  {Sylju\aa{}sen}}, \bibinfo {author} {\bibfnamefont {B.~L.}\ \bibnamefont
  {Altshuler}},\ and\ \bibinfo {author} {\bibfnamefont {V.}~\bibnamefont
  {Fal'ko}},\ }\bibfield  {title} {\bibinfo {title} {Ordered states of adatoms
  on graphene},\ }\href {https://doi.org/10.1103/PhysRevB.80.233409} {\bibfield
   {journal} {\bibinfo  {journal} {Phys. Rev. B}\ }\textbf {\bibinfo {volume}
  {80}},\ \bibinfo {pages} {233409} (\bibinfo {year} {2009})}\BibitemShut
  {NoStop}%
\bibitem [{\citenamefont {Ryu}\ \emph {et~al.}(2009)\citenamefont {Ryu},
  \citenamefont {Mudry}, \citenamefont {Hou},\ and\ \citenamefont
  {Chamon}}]{PhysRevB.80.205319}%
  \BibitemOpen
  \bibfield  {author} {\bibinfo {author} {\bibfnamefont {S.}~\bibnamefont
  {Ryu}}, \bibinfo {author} {\bibfnamefont {C.}~\bibnamefont {Mudry}}, \bibinfo
  {author} {\bibfnamefont {C.-Y.}\ \bibnamefont {Hou}},\ and\ \bibinfo {author}
  {\bibfnamefont {C.}~\bibnamefont {Chamon}},\ }\bibfield  {title} {\bibinfo
  {title} {Masses in graphenelike two-dimensional electronic systems:
  Topological defects in order parameters and their fractional exchange
  statistics},\ }\href {https://doi.org/10.1103/PhysRevB.80.205319} {\bibfield
  {journal} {\bibinfo  {journal} {Phys. Rev. B}\ }\textbf {\bibinfo {volume}
  {80}},\ \bibinfo {pages} {205319} (\bibinfo {year} {2009})}\BibitemShut
  {NoStop}%
\bibitem [{\citenamefont {Weeks}\ and\ \citenamefont
  {Franz}(2010)}]{PhysRevB.81.085105}%
  \BibitemOpen
  \bibfield  {author} {\bibinfo {author} {\bibfnamefont {C.}~\bibnamefont
  {Weeks}}\ and\ \bibinfo {author} {\bibfnamefont {M.}~\bibnamefont {Franz}},\
  }\bibfield  {title} {\bibinfo {title} {Interaction-driven instabilities of a
  dirac semimetal},\ }\href {https://doi.org/10.1103/PhysRevB.81.085105}
  {\bibfield  {journal} {\bibinfo  {journal} {Phys. Rev. B}\ }\textbf {\bibinfo
  {volume} {81}},\ \bibinfo {pages} {085105} (\bibinfo {year}
  {2010})}\BibitemShut {NoStop}%
\bibitem [{\citenamefont {Kopylov}\ \emph {et~al.}(2011)\citenamefont
  {Kopylov}, \citenamefont {Cheianov}, \citenamefont {Altshuler},\ and\
  \citenamefont {Fal'ko}}]{PhysRevB.83.201401}%
  \BibitemOpen
  \bibfield  {author} {\bibinfo {author} {\bibfnamefont {S.}~\bibnamefont
  {Kopylov}}, \bibinfo {author} {\bibfnamefont {V.}~\bibnamefont {Cheianov}},
  \bibinfo {author} {\bibfnamefont {B.~L.}\ \bibnamefont {Altshuler}},\ and\
  \bibinfo {author} {\bibfnamefont {V.~I.}\ \bibnamefont {Fal'ko}},\ }\bibfield
   {title} {\bibinfo {title} {Transport anomaly at the ordering transition for
  adatoms on graphene},\ }\href {https://doi.org/10.1103/PhysRevB.83.201401}
  {\bibfield  {journal} {\bibinfo  {journal} {Phys. Rev. B}\ }\textbf {\bibinfo
  {volume} {83}},\ \bibinfo {pages} {201401} (\bibinfo {year}
  {2011})}\BibitemShut {NoStop}%
\bibitem [{\citenamefont {Garc\'{\i}a-Mart\'{\i}nez}\ \emph
  {et~al.}(2013)\citenamefont {Garc\'{\i}a-Mart\'{\i}nez}, \citenamefont
  {Grushin}, \citenamefont {Neupert}, \citenamefont {Valenzuela},\ and\
  \citenamefont {Castro}}]{PhysRevB.88.245123}%
  \BibitemOpen
  \bibfield  {author} {\bibinfo {author} {\bibfnamefont {N.~A.}\ \bibnamefont
  {Garc\'{\i}a-Mart\'{\i}nez}}, \bibinfo {author} {\bibfnamefont {A.~G.}\
  \bibnamefont {Grushin}}, \bibinfo {author} {\bibfnamefont {T.}~\bibnamefont
  {Neupert}}, \bibinfo {author} {\bibfnamefont {B.}~\bibnamefont
  {Valenzuela}},\ and\ \bibinfo {author} {\bibfnamefont {E.~V.}\ \bibnamefont
  {Castro}},\ }\bibfield  {title} {\bibinfo {title} {Interaction-driven phases
  in the half-filled spinless honeycomb lattice from exact diagonalization},\
  }\href {https://doi.org/10.1103/PhysRevB.88.245123} {\bibfield  {journal}
  {\bibinfo  {journal} {Phys. Rev. B}\ }\textbf {\bibinfo {volume} {88}},\
  \bibinfo {pages} {245123} (\bibinfo {year} {2013})}\BibitemShut {NoStop}%
\bibitem [{\citenamefont {Mu}\ \emph {et~al.}(2022)\citenamefont {Mu},
  \citenamefont {Liu}, \citenamefont {Hu},\ and\ \citenamefont
  {Wang}}]{Mu2022}%
  \BibitemOpen
  \bibfield  {author} {\bibinfo {author} {\bibfnamefont {H.}~\bibnamefont
  {Mu}}, \bibinfo {author} {\bibfnamefont {B.}~\bibnamefont {Liu}}, \bibinfo
  {author} {\bibfnamefont {T.}~\bibnamefont {Hu}},\ and\ \bibinfo {author}
  {\bibfnamefont {Z.}~\bibnamefont {Wang}},\ }\bibfield  {title} {\bibinfo
  {title} {Kekul{\'e} lattice in graphdiyne: Coexistence of phononic and
  electronic second-order topological insulator},\ }\href
  {https://doi.org/10.1021/acs.nanolett.1c04239} {\bibfield  {journal}
  {\bibinfo  {journal} {Nano Lett.}\ }\textbf {\bibinfo {volume} {22}},\
  \bibinfo {pages} {1122} (\bibinfo {year} {2022})}\BibitemShut {NoStop}%
\bibitem [{\citenamefont {Guti{\'e}rrez}\ \emph {et~al.}(2016)\citenamefont
  {Guti{\'e}rrez}, \citenamefont {Kim}, \citenamefont {Brown}, \citenamefont
  {Schiros}, \citenamefont {Nordlund}, \citenamefont {Lochocki}, \citenamefont
  {Shen}, \citenamefont {Park},\ and\ \citenamefont
  {Pasupathy}}]{Gutierrez2016}%
  \BibitemOpen
  \bibfield  {author} {\bibinfo {author} {\bibfnamefont {C.}~\bibnamefont
  {Guti{\'e}rrez}}, \bibinfo {author} {\bibfnamefont {C.-J.}\ \bibnamefont
  {Kim}}, \bibinfo {author} {\bibfnamefont {L.}~\bibnamefont {Brown}}, \bibinfo
  {author} {\bibfnamefont {T.}~\bibnamefont {Schiros}}, \bibinfo {author}
  {\bibfnamefont {D.}~\bibnamefont {Nordlund}}, \bibinfo {author}
  {\bibfnamefont {E.}~\bibnamefont {Lochocki}}, \bibinfo {author}
  {\bibfnamefont {K.~M.}\ \bibnamefont {Shen}}, \bibinfo {author}
  {\bibfnamefont {J.}~\bibnamefont {Park}},\ and\ \bibinfo {author}
  {\bibfnamefont {A.~N.}\ \bibnamefont {Pasupathy}},\ }\bibfield  {title}
  {\bibinfo {title} {Imaging chiral symmetry breaking from kekul{\'e} bond
  order in graphene},\ }\href {https://doi.org/10.1038/nphys3776} {\bibfield
  {journal} {\bibinfo  {journal} {Nat. Phys.}\ }\textbf {\bibinfo {volume}
  {12}},\ \bibinfo {pages} {950} (\bibinfo {year} {2016})}\BibitemShut
  {NoStop}%
\bibitem [{\citenamefont {Liu}\ \emph {et~al.}(2017)\citenamefont {Liu},
  \citenamefont {Lian}, \citenamefont {Li}, \citenamefont {Xu},\ and\
  \citenamefont {Duan}}]{PhysRevLett.119.255901}%
  \BibitemOpen
  \bibfield  {author} {\bibinfo {author} {\bibfnamefont {Y.}~\bibnamefont
  {Liu}}, \bibinfo {author} {\bibfnamefont {C.-S.}\ \bibnamefont {Lian}},
  \bibinfo {author} {\bibfnamefont {Y.}~\bibnamefont {Li}}, \bibinfo {author}
  {\bibfnamefont {Y.}~\bibnamefont {Xu}},\ and\ \bibinfo {author}
  {\bibfnamefont {W.}~\bibnamefont {Duan}},\ }\bibfield  {title} {\bibinfo
  {title} {Pseudospins and topological effects of phonons in a kekul\'e
  lattice},\ }\href {https://doi.org/10.1103/PhysRevLett.119.255901} {\bibfield
   {journal} {\bibinfo  {journal} {Phys. Rev. Lett.}\ }\textbf {\bibinfo
  {volume} {119}},\ \bibinfo {pages} {255901} (\bibinfo {year}
  {2017})}\BibitemShut {NoStop}%
\bibitem [{\citenamefont {Bao}\ \emph {et~al.}(2021)\citenamefont {Bao},
  \citenamefont {Zhang}, \citenamefont {Zhang}, \citenamefont {Wu},
  \citenamefont {Luo}, \citenamefont {Zhou}, \citenamefont {Li}, \citenamefont
  {Hou}, \citenamefont {Yao}, \citenamefont {Liu}, \citenamefont {Yu},
  \citenamefont {Li}, \citenamefont {Duan}, \citenamefont {Yao}, \citenamefont
  {Wang},\ and\ \citenamefont {Zhou}}]{PhysRevLett.126.206804}%
  \BibitemOpen
  \bibfield  {author} {\bibinfo {author} {\bibfnamefont {C.}~\bibnamefont
  {Bao}}, \bibinfo {author} {\bibfnamefont {H.}~\bibnamefont {Zhang}}, \bibinfo
  {author} {\bibfnamefont {T.}~\bibnamefont {Zhang}}, \bibinfo {author}
  {\bibfnamefont {X.}~\bibnamefont {Wu}}, \bibinfo {author} {\bibfnamefont
  {L.}~\bibnamefont {Luo}}, \bibinfo {author} {\bibfnamefont {S.}~\bibnamefont
  {Zhou}}, \bibinfo {author} {\bibfnamefont {Q.}~\bibnamefont {Li}}, \bibinfo
  {author} {\bibfnamefont {Y.}~\bibnamefont {Hou}}, \bibinfo {author}
  {\bibfnamefont {W.}~\bibnamefont {Yao}}, \bibinfo {author} {\bibfnamefont
  {L.}~\bibnamefont {Liu}}, \bibinfo {author} {\bibfnamefont {P.}~\bibnamefont
  {Yu}}, \bibinfo {author} {\bibfnamefont {J.}~\bibnamefont {Li}}, \bibinfo
  {author} {\bibfnamefont {W.}~\bibnamefont {Duan}}, \bibinfo {author}
  {\bibfnamefont {H.}~\bibnamefont {Yao}}, \bibinfo {author} {\bibfnamefont
  {Y.}~\bibnamefont {Wang}},\ and\ \bibinfo {author} {\bibfnamefont
  {S.}~\bibnamefont {Zhou}},\ }\bibfield  {title} {\bibinfo {title}
  {Experimental evidence of chiral symmetry breaking in kekul\'e-ordered
  graphene},\ }\href {https://doi.org/10.1103/PhysRevLett.126.206804}
  {\bibfield  {journal} {\bibinfo  {journal} {Phys. Rev. Lett.}\ }\textbf
  {\bibinfo {volume} {126}},\ \bibinfo {pages} {206804} (\bibinfo {year}
  {2021})}\BibitemShut {NoStop}%
\bibitem [{\citenamefont {Scheer}\ and\ \citenamefont
  {Lian}(2023)}]{PhysRevLett.131.266501}%
  \BibitemOpen
  \bibfield  {author} {\bibinfo {author} {\bibfnamefont {M.~G.}\ \bibnamefont
  {Scheer}}\ and\ \bibinfo {author} {\bibfnamefont {B.}~\bibnamefont {Lian}},\
  }\bibfield  {title} {\bibinfo {title} {Twistronics of kekul\'e graphene:
  Honeycomb and kagome flat bands},\ }\href
  {https://doi.org/10.1103/PhysRevLett.131.266501} {\bibfield  {journal}
  {\bibinfo  {journal} {Phys. Rev. Lett.}\ }\textbf {\bibinfo {volume} {131}},\
  \bibinfo {pages} {266501} (\bibinfo {year} {2023})}\BibitemShut {NoStop}%
\bibitem [{\citenamefont {Kawarabayashi}\ \emph {et~al.}(2021)\citenamefont
  {Kawarabayashi}, \citenamefont {Inoue}, \citenamefont {Itagaki},
  \citenamefont {Hatsugai},\ and\ \citenamefont
  {Aoki}}]{KAWARABAYASHI2021168440}%
  \BibitemOpen
  \bibfield  {author} {\bibinfo {author} {\bibfnamefont {T.}~\bibnamefont
  {Kawarabayashi}}, \bibinfo {author} {\bibfnamefont {Y.}~\bibnamefont
  {Inoue}}, \bibinfo {author} {\bibfnamefont {R.}~\bibnamefont {Itagaki}},
  \bibinfo {author} {\bibfnamefont {Y.}~\bibnamefont {Hatsugai}},\ and\
  \bibinfo {author} {\bibfnamefont {H.}~\bibnamefont {Aoki}},\ }\bibfield
  {title} {\bibinfo {title} {Robust zero modes in disordered two-dimensional
  honeycomb lattice with kekulé bond ordering},\ }\href
  {https://doi.org/https://doi.org/10.1016/j.aop.2021.168440} {\bibfield
  {journal} {\bibinfo  {journal} {Annal. Phys.}\ }\textbf {\bibinfo {volume}
  {435}},\ \bibinfo {pages} {168440} (\bibinfo {year} {2021})},\ \bibinfo
  {note} {special Issue on Localisation 2020}\BibitemShut {NoStop}%
\bibitem [{\citenamefont {Otsuka}\ and\ \citenamefont
  {Yunoki}(2024)}]{PhysRevB.109.115131}%
  \BibitemOpen
  \bibfield  {author} {\bibinfo {author} {\bibfnamefont {Y.}~\bibnamefont
  {Otsuka}}\ and\ \bibinfo {author} {\bibfnamefont {S.}~\bibnamefont
  {Yunoki}},\ }\bibfield  {title} {\bibinfo {title} {Kekul\'e valence bond
  order in the hubbard model on the honeycomb lattice with possible lattice
  distortions for graphene},\ }\href
  {https://doi.org/10.1103/PhysRevB.109.115131} {\bibfield  {journal} {\bibinfo
   {journal} {Phys. Rev. B}\ }\textbf {\bibinfo {volume} {109}},\ \bibinfo
  {pages} {115131} (\bibinfo {year} {2024})}\BibitemShut {NoStop}%
\bibitem [{\citenamefont {Andrade}\ \emph {et~al.}(2020)\citenamefont
  {Andrade}, \citenamefont {Carrillo-Bastos}, \citenamefont {Pantaleón},\ and\
  \citenamefont {Mireles}}]{10.1063/1.5133091}%
  \BibitemOpen
  \bibfield  {author} {\bibinfo {author} {\bibfnamefont {E.}~\bibnamefont
  {Andrade}}, \bibinfo {author} {\bibfnamefont {R.}~\bibnamefont
  {Carrillo-Bastos}}, \bibinfo {author} {\bibfnamefont {P.~A.}\ \bibnamefont
  {Pantaleón}},\ and\ \bibinfo {author} {\bibfnamefont {F.}~\bibnamefont
  {Mireles}},\ }\bibfield  {title} {\bibinfo {title} {{Resonant transport in
  Kekulé-distorted graphene nanoribbons}},\ }\href
  {https://doi.org/10.1063/1.5133091} {\bibfield  {journal} {\bibinfo
  {journal} {J. Appl. Phys.}\ }\textbf {\bibinfo {volume} {127}},\ \bibinfo
  {pages} {054304} (\bibinfo {year} {2020})}\BibitemShut {NoStop}%
\bibitem [{\citenamefont {Freeney}\ \emph {et~al.}(2020)\citenamefont
  {Freeney}, \citenamefont {van~den Broeke}, \citenamefont {Harsveld van~der
  Veen}, \citenamefont {Swart},\ and\ \citenamefont
  {Morais~Smith}}]{PhysRevLett.124.236404}%
  \BibitemOpen
  \bibfield  {author} {\bibinfo {author} {\bibfnamefont {S.~E.}\ \bibnamefont
  {Freeney}}, \bibinfo {author} {\bibfnamefont {J.~J.}\ \bibnamefont {van~den
  Broeke}}, \bibinfo {author} {\bibfnamefont {A.~J.~J.}\ \bibnamefont {Harsveld
  van~der Veen}}, \bibinfo {author} {\bibfnamefont {I.}~\bibnamefont {Swart}},\
  and\ \bibinfo {author} {\bibfnamefont {C.}~\bibnamefont {Morais~Smith}},\
  }\bibfield  {title} {\bibinfo {title} {Edge-dependent topology in kekul\'e
  lattices},\ }\href {https://doi.org/10.1103/PhysRevLett.124.236404}
  {\bibfield  {journal} {\bibinfo  {journal} {Phys. Rev. Lett.}\ }\textbf
  {\bibinfo {volume} {124}},\ \bibinfo {pages} {236404} (\bibinfo {year}
  {2020})}\BibitemShut {NoStop}%
\bibitem [{\citenamefont {Jiang}\ \emph {et~al.}(2024)\citenamefont {Jiang},
  \citenamefont {Kariyado},\ and\ \citenamefont
  {Hu}}]{doi:10.7566/JPSJ.93.033703}%
  \BibitemOpen
  \bibfield  {author} {\bibinfo {author} {\bibfnamefont {Y.-C.}\ \bibnamefont
  {Jiang}}, \bibinfo {author} {\bibfnamefont {T.}~\bibnamefont {Kariyado}},\
  and\ \bibinfo {author} {\bibfnamefont {X.}~\bibnamefont {Hu}},\ }\bibfield
  {title} {\bibinfo {title} {Higher-order topology in honeycomb lattice with
  y-kekulé distortions},\ }\href {https://doi.org/10.7566/JPSJ.93.033703}
  {\bibfield  {journal} {\bibinfo  {journal} {J. Phys. Soc. Jpn.}\ }\textbf
  {\bibinfo {volume} {93}},\ \bibinfo {pages} {033703} (\bibinfo {year}
  {2024})}\BibitemShut {NoStop}%
\bibitem [{\citenamefont {Zhang}\ \emph {et~al.}(2023)\citenamefont {Zhang},
  \citenamefont {Wang}, \citenamefont {Li}, \citenamefont {Zhai},\ and\
  \citenamefont {Song}}]{Zhang_2023}%
  \BibitemOpen
  \bibfield  {author} {\bibinfo {author} {\bibfnamefont {P.}~\bibnamefont
  {Zhang}}, \bibinfo {author} {\bibfnamefont {C.}~\bibnamefont {Wang}},
  \bibinfo {author} {\bibfnamefont {Y.-X.}\ \bibnamefont {Li}}, \bibinfo
  {author} {\bibfnamefont {L.}~\bibnamefont {Zhai}},\ and\ \bibinfo {author}
  {\bibfnamefont {J.}~\bibnamefont {Song}},\ }\bibfield  {title} {\bibinfo
  {title} {The transport properties of kekulé-ordered graphene p–n
  junction},\ }\href {https://doi.org/10.1088/1367-2630/ad091d} {\bibfield
  {journal} {\bibinfo  {journal} {New J. Phys.}\ }\textbf {\bibinfo {volume}
  {25}},\ \bibinfo {pages} {113021} (\bibinfo {year} {2023})}\BibitemShut
  {NoStop}%
\bibitem [{\citenamefont {Mohammadi}(2021)}]{Mohammadi_2021}%
  \BibitemOpen
  \bibfield  {author} {\bibinfo {author} {\bibfnamefont {Y.}~\bibnamefont
  {Mohammadi}},\ }\bibfield  {title} {\bibinfo {title} {Magneto-optical
  conductivity of graphene: Signatures of a uniform y-shaped kekule lattice
  distortion},\ }\href {https://doi.org/10.1149/2162-8777/ac08d5} {\bibfield
  {journal} {\bibinfo  {journal} {ECS J. Solid State Sci. Technol.}\ }\textbf
  {\bibinfo {volume} {10}},\ \bibinfo {pages} {061011} (\bibinfo {year}
  {2021})}\BibitemShut {NoStop}%
\bibitem [{\citenamefont {Gamayun}\ \emph {et~al.}(2018)\citenamefont
  {Gamayun}, \citenamefont {Ostroukh}, \citenamefont {Gnezdilov}, \citenamefont
  {Adagideli},\ and\ \citenamefont {Beenakker}}]{Gamayun_2018}%
  \BibitemOpen
  \bibfield  {author} {\bibinfo {author} {\bibfnamefont {O.~V.}\ \bibnamefont
  {Gamayun}}, \bibinfo {author} {\bibfnamefont {V.~P.}\ \bibnamefont
  {Ostroukh}}, \bibinfo {author} {\bibfnamefont {N.~V.}\ \bibnamefont
  {Gnezdilov}}, \bibinfo {author} {\bibfnamefont {I.}~\bibnamefont
  {Adagideli}},\ and\ \bibinfo {author} {\bibfnamefont {C.~W.~J.}\ \bibnamefont
  {Beenakker}},\ }\bibfield  {title} {\bibinfo {title} {Valley-momentum locking
  in a graphene superlattice with y-shaped kekulé bond texture},\ }\href
  {https://doi.org/10.1088/1367-2630/aaa7e5} {\bibfield  {journal} {\bibinfo
  {journal} {New J. Phys.}\ }\textbf {\bibinfo {volume} {20}},\ \bibinfo
  {pages} {023016} (\bibinfo {year} {2018})}\BibitemShut {NoStop}%
\bibitem [{\citenamefont {Gao}\ \emph {et~al.}(2020)\citenamefont {Gao},
  \citenamefont {Yang}, \citenamefont {Lin}, \citenamefont {Zhang},
  \citenamefont {Li}, \citenamefont {Bo}, \citenamefont {Wang},\ and\
  \citenamefont {Lu}}]{Gao2020}%
  \BibitemOpen
  \bibfield  {author} {\bibinfo {author} {\bibfnamefont {X.}~\bibnamefont
  {Gao}}, \bibinfo {author} {\bibfnamefont {L.}~\bibnamefont {Yang}}, \bibinfo
  {author} {\bibfnamefont {H.}~\bibnamefont {Lin}}, \bibinfo {author}
  {\bibfnamefont {L.}~\bibnamefont {Zhang}}, \bibinfo {author} {\bibfnamefont
  {J.}~\bibnamefont {Li}}, \bibinfo {author} {\bibfnamefont {F.}~\bibnamefont
  {Bo}}, \bibinfo {author} {\bibfnamefont {Z.}~\bibnamefont {Wang}},\ and\
  \bibinfo {author} {\bibfnamefont {L.}~\bibnamefont {Lu}},\ }\bibfield
  {title} {\bibinfo {title} {Dirac-vortex topological cavities},\ }\href
  {https://doi.org/10.1038/s41565-020-0773-7} {\bibfield  {journal} {\bibinfo
  {journal} {Nat. Nanotechnol.}\ }\textbf {\bibinfo {volume} {15}},\ \bibinfo
  {pages} {1012} (\bibinfo {year} {2020})}\BibitemShut {NoStop}%
\bibitem [{\citenamefont {Wei}\ \emph {et~al.}(2022)\citenamefont {Wei},
  \citenamefont {Liu}, \citenamefont {Wang}, \citenamefont {Song},\ and\
  \citenamefont {Xiao}}]{Wei:22}%
  \BibitemOpen
  \bibfield  {author} {\bibinfo {author} {\bibfnamefont {G.}~\bibnamefont
  {Wei}}, \bibinfo {author} {\bibfnamefont {Z.}~\bibnamefont {Liu}}, \bibinfo
  {author} {\bibfnamefont {L.}~\bibnamefont {Wang}}, \bibinfo {author}
  {\bibfnamefont {J.}~\bibnamefont {Song}},\ and\ \bibinfo {author}
  {\bibfnamefont {J.-J.}\ \bibnamefont {Xiao}},\ }\bibfield  {title} {\bibinfo
  {title} {Coexisting valley and pseudo-spin topological edge states in
  photonic topological insulators made of distorted kekulé lattices},\ }\href
  {https://doi.org/10.1364/PRJ.453803} {\bibfield  {journal} {\bibinfo
  {journal} {Photon. Res.}\ }\textbf {\bibinfo {volume} {10}},\ \bibinfo
  {pages} {999} (\bibinfo {year} {2022})}\BibitemShut {NoStop}%
\bibitem [{\citenamefont {Huang}\ \emph {et~al.}(2023)\citenamefont {Huang},
  \citenamefont {Chen}, \citenamefont {Mao},\ and\ \citenamefont
  {Wang}}]{Huang_2024}%
  \BibitemOpen
  \bibfield  {author} {\bibinfo {author} {\bibfnamefont {H.}~\bibnamefont
  {Huang}}, \bibinfo {author} {\bibfnamefont {J.}~\bibnamefont {Chen}},
  \bibinfo {author} {\bibfnamefont {L.}~\bibnamefont {Mao}},\ and\ \bibinfo
  {author} {\bibfnamefont {R.}~\bibnamefont {Wang}},\ }\bibfield  {title}
  {\bibinfo {title} {Simultaneous pseudospin and valley topological edge states
  of elastic waves in phononic crystals made of distorted kekulé lattices},\
  }\href {https://doi.org/10.1088/1361-648X/ad162e} {\bibfield  {journal}
  {\bibinfo  {journal} {J. Phys. Condens. Matter}\ }\textbf {\bibinfo {volume}
  {36}},\ \bibinfo {pages} {135402} (\bibinfo {year} {2023})}\BibitemShut
  {NoStop}%
\bibitem [{\citenamefont {Xie}\ \emph {et~al.}(2019)\citenamefont {Xie},
  \citenamefont {Liu}, \citenamefont {Cheng}, \citenamefont {Liu},
  \citenamefont {Chen},\ and\ \citenamefont {Tian}}]{PhysRevApplied.11.044086}%
  \BibitemOpen
  \bibfield  {author} {\bibinfo {author} {\bibfnamefont {B.}~\bibnamefont
  {Xie}}, \bibinfo {author} {\bibfnamefont {H.}~\bibnamefont {Liu}}, \bibinfo
  {author} {\bibfnamefont {H.}~\bibnamefont {Cheng}}, \bibinfo {author}
  {\bibfnamefont {Z.}~\bibnamefont {Liu}}, \bibinfo {author} {\bibfnamefont
  {S.}~\bibnamefont {Chen}},\ and\ \bibinfo {author} {\bibfnamefont
  {J.}~\bibnamefont {Tian}},\ }\bibfield  {title} {\bibinfo {title} {Acoustic
  topological transport and refraction in a kekul\'e lattice},\ }\href
  {https://doi.org/10.1103/PhysRevApplied.11.044086} {\bibfield  {journal}
  {\bibinfo  {journal} {Phys. Rev. Appl.}\ }\textbf {\bibinfo {volume} {11}},\
  \bibinfo {pages} {044086} (\bibinfo {year} {2019})}\BibitemShut {NoStop}%
\bibitem [{\citenamefont {Ma}\ \emph {et~al.}(2018)\citenamefont {Ma},
  \citenamefont {Fu}, \citenamefont {Sui}, \citenamefont {Bai}, \citenamefont
  {Qiao}, \citenamefont {Yan}, \citenamefont {Zhang}, \citenamefont {Hu},
  \citenamefont {Xiao}, \citenamefont {Mao}, \citenamefont {Duan},\ and\
  \citenamefont {He}}]{Ma2018}%
  \BibitemOpen
  \bibfield  {author} {\bibinfo {author} {\bibfnamefont {D.}~\bibnamefont
  {Ma}}, \bibinfo {author} {\bibfnamefont {Z.}~\bibnamefont {Fu}}, \bibinfo
  {author} {\bibfnamefont {X.}~\bibnamefont {Sui}}, \bibinfo {author}
  {\bibfnamefont {K.}~\bibnamefont {Bai}}, \bibinfo {author} {\bibfnamefont
  {J.}~\bibnamefont {Qiao}}, \bibinfo {author} {\bibfnamefont {C.}~\bibnamefont
  {Yan}}, \bibinfo {author} {\bibfnamefont {Y.}~\bibnamefont {Zhang}}, \bibinfo
  {author} {\bibfnamefont {J.}~\bibnamefont {Hu}}, \bibinfo {author}
  {\bibfnamefont {Q.}~\bibnamefont {Xiao}}, \bibinfo {author} {\bibfnamefont
  {X.}~\bibnamefont {Mao}}, \bibinfo {author} {\bibfnamefont {W.}~\bibnamefont
  {Duan}},\ and\ \bibinfo {author} {\bibfnamefont {L.}~\bibnamefont {He}},\
  }\bibfield  {title} {\bibinfo {title} {Modulating the electronic properties
  of graphene by self-organized sulfur identical nanoclusters and atomic
  superlattices confined at an interface},\ }\href
  {https://doi.org/10.1021/acsnano.8b04874} {\bibfield  {journal} {\bibinfo
  {journal} {ACS Nano}\ }\textbf {\bibinfo {volume} {12}},\ \bibinfo {pages}
  {10984} (\bibinfo {year} {2018})}\BibitemShut {NoStop}%
\bibitem [{\citenamefont {Essmann}\ and\ \citenamefont
  {Träuble}(1967)}]{ESSMANN1967526}%
  \BibitemOpen
  \bibfield  {author} {\bibinfo {author} {\bibfnamefont {U.}~\bibnamefont
  {Essmann}}\ and\ \bibinfo {author} {\bibfnamefont {H.}~\bibnamefont
  {Träuble}},\ }\bibfield  {title} {\bibinfo {title} {The direct observation
  of individual flux lines in type ii superconductors},\ }\href
  {https://doi.org/https://doi.org/10.1016/0375-9601(67)90819-5} {\bibfield
  {journal} {\bibinfo  {journal} {Phys. Lett. A}\ }\textbf {\bibinfo {volume}
  {24}},\ \bibinfo {pages} {526} (\bibinfo {year} {1967})}\BibitemShut
  {NoStop}%
\bibitem [{\citenamefont {Mühlbauer}\ \emph {et~al.}(2009)\citenamefont
  {Mühlbauer}, \citenamefont {Binz}, \citenamefont {Jonietz}, \citenamefont
  {Pfleiderer}, \citenamefont {Rosch}, \citenamefont {Neubauer}, \citenamefont
  {Georgii},\ and\ \citenamefont {Böni}}]{doi:10.1126/science.1166767}%
  \BibitemOpen
  \bibfield  {author} {\bibinfo {author} {\bibfnamefont {S.}~\bibnamefont
  {Mühlbauer}}, \bibinfo {author} {\bibfnamefont {B.}~\bibnamefont {Binz}},
  \bibinfo {author} {\bibfnamefont {F.}~\bibnamefont {Jonietz}}, \bibinfo
  {author} {\bibfnamefont {C.}~\bibnamefont {Pfleiderer}}, \bibinfo {author}
  {\bibfnamefont {A.}~\bibnamefont {Rosch}}, \bibinfo {author} {\bibfnamefont
  {A.}~\bibnamefont {Neubauer}}, \bibinfo {author} {\bibfnamefont
  {R.}~\bibnamefont {Georgii}},\ and\ \bibinfo {author} {\bibfnamefont
  {P.}~\bibnamefont {Böni}},\ }\bibfield  {title} {\bibinfo {title} {Skyrmion
  lattice in a chiral magnet},\ }\href
  {https://doi.org/10.1126/science.1166767} {\bibfield  {journal} {\bibinfo
  {journal} {Science}\ }\textbf {\bibinfo {volume} {323}},\ \bibinfo {pages}
  {915} (\bibinfo {year} {2009})}\BibitemShut {NoStop}%
\bibitem [{\citenamefont {Kim}\ \emph {et~al.}(2024)\citenamefont {Kim},
  \citenamefont {Go}, \citenamefont {Park},\ and\ \citenamefont
  {Kim}}]{Kim2024}%
  \BibitemOpen
  \bibfield  {author} {\bibinfo {author} {\bibfnamefont {K.-M.}\ \bibnamefont
  {Kim}}, \bibinfo {author} {\bibfnamefont {G.}~\bibnamefont {Go}}, \bibinfo
  {author} {\bibfnamefont {M.~J.}\ \bibnamefont {Park}},\ and\ \bibinfo
  {author} {\bibfnamefont {S.~K.}\ \bibnamefont {Kim}},\ }\bibfield  {title}
  {\bibinfo {title} {Emergence of stable meron quartets in twisted magnets},\
  }\href {https://doi.org/10.1021/acs.nanolett.3c03246} {\bibfield  {journal}
  {\bibinfo  {journal} {Nano Lett.}\ }\textbf {\bibinfo {volume} {24}},\
  \bibinfo {pages} {74} (\bibinfo {year} {2024})}\BibitemShut {NoStop}%
\bibitem [{\citenamefont {Yu}\ \emph {et~al.}(2012)\citenamefont {Yu},
  \citenamefont {Kanazawa}, \citenamefont {Zhang}, \citenamefont {Nagai},
  \citenamefont {Hara}, \citenamefont {Kimoto}, \citenamefont {Matsui},
  \citenamefont {Onose},\ and\ \citenamefont {Tokura}}]{Yu2012}%
  \BibitemOpen
  \bibfield  {author} {\bibinfo {author} {\bibfnamefont {X.~Z.}\ \bibnamefont
  {Yu}}, \bibinfo {author} {\bibfnamefont {N.}~\bibnamefont {Kanazawa}},
  \bibinfo {author} {\bibfnamefont {W.~Z.}\ \bibnamefont {Zhang}}, \bibinfo
  {author} {\bibfnamefont {T.}~\bibnamefont {Nagai}}, \bibinfo {author}
  {\bibfnamefont {T.}~\bibnamefont {Hara}}, \bibinfo {author} {\bibfnamefont
  {K.}~\bibnamefont {Kimoto}}, \bibinfo {author} {\bibfnamefont
  {Y.}~\bibnamefont {Matsui}}, \bibinfo {author} {\bibfnamefont
  {Y.}~\bibnamefont {Onose}},\ and\ \bibinfo {author} {\bibfnamefont
  {Y.}~\bibnamefont {Tokura}},\ }\bibfield  {title} {\bibinfo {title} {Skyrmion
  flow near room temperature in an ultralow current density},\ }\href
  {https://doi.org/10.1038/ncomms1990} {\bibfield  {journal} {\bibinfo
  {journal} {Nat. Commun.}\ }\textbf {\bibinfo {volume} {3}},\ \bibinfo {pages}
  {988} (\bibinfo {year} {2012})}\BibitemShut {NoStop}%
\bibitem [{\citenamefont {Yu}\ \emph {et~al.}(2018)\citenamefont {Yu},
  \citenamefont {Koshibae}, \citenamefont {Tokunaga}, \citenamefont {Shibata},
  \citenamefont {Taguchi}, \citenamefont {Nagaosa},\ and\ \citenamefont
  {Tokura}}]{Yu2018}%
  \BibitemOpen
  \bibfield  {author} {\bibinfo {author} {\bibfnamefont {X.~Z.}\ \bibnamefont
  {Yu}}, \bibinfo {author} {\bibfnamefont {W.}~\bibnamefont {Koshibae}},
  \bibinfo {author} {\bibfnamefont {Y.}~\bibnamefont {Tokunaga}}, \bibinfo
  {author} {\bibfnamefont {K.}~\bibnamefont {Shibata}}, \bibinfo {author}
  {\bibfnamefont {Y.}~\bibnamefont {Taguchi}}, \bibinfo {author} {\bibfnamefont
  {N.}~\bibnamefont {Nagaosa}},\ and\ \bibinfo {author} {\bibfnamefont
  {Y.}~\bibnamefont {Tokura}},\ }\bibfield  {title} {\bibinfo {title}
  {Transformation between meron and skyrmion topological spin textures in a
  chiral magnet},\ }\href {https://doi.org/10.1038/s41586-018-0745-3}
  {\bibfield  {journal} {\bibinfo  {journal} {Nature}\ }\textbf {\bibinfo
  {volume} {564}},\ \bibinfo {pages} {95} (\bibinfo {year} {2018})}\BibitemShut
  {NoStop}%
\bibitem [{\citenamefont {Zhang}\ \emph {et~al.}(2022)\citenamefont {Zhang},
  \citenamefont {Raftrey}, \citenamefont {Chan}, \citenamefont {Shao},
  \citenamefont {Chen}, \citenamefont {Chen}, \citenamefont {Huang},
  \citenamefont {Reichanadter}, \citenamefont {Dong}, \citenamefont {Susarla},
  \citenamefont {Caretta}, \citenamefont {Chen}, \citenamefont {Yao},
  \citenamefont {Fischer}, \citenamefont {Neaton}, \citenamefont {Wu},
  \citenamefont {Muller}, \citenamefont {Birgeneau},\ and\ \citenamefont
  {Ramesh}}]{doi:10.1126/sciadv.abm7103}%
  \BibitemOpen
  \bibfield  {author} {\bibinfo {author} {\bibfnamefont {H.}~\bibnamefont
  {Zhang}}, \bibinfo {author} {\bibfnamefont {D.}~\bibnamefont {Raftrey}},
  \bibinfo {author} {\bibfnamefont {Y.-T.}\ \bibnamefont {Chan}}, \bibinfo
  {author} {\bibfnamefont {Y.-T.}\ \bibnamefont {Shao}}, \bibinfo {author}
  {\bibfnamefont {R.}~\bibnamefont {Chen}}, \bibinfo {author} {\bibfnamefont
  {X.}~\bibnamefont {Chen}}, \bibinfo {author} {\bibfnamefont {X.}~\bibnamefont
  {Huang}}, \bibinfo {author} {\bibfnamefont {J.~T.}\ \bibnamefont
  {Reichanadter}}, \bibinfo {author} {\bibfnamefont {K.}~\bibnamefont {Dong}},
  \bibinfo {author} {\bibfnamefont {S.}~\bibnamefont {Susarla}}, \bibinfo
  {author} {\bibfnamefont {L.}~\bibnamefont {Caretta}}, \bibinfo {author}
  {\bibfnamefont {Z.}~\bibnamefont {Chen}}, \bibinfo {author} {\bibfnamefont
  {J.}~\bibnamefont {Yao}}, \bibinfo {author} {\bibfnamefont {P.}~\bibnamefont
  {Fischer}}, \bibinfo {author} {\bibfnamefont {J.~B.}\ \bibnamefont {Neaton}},
  \bibinfo {author} {\bibfnamefont {W.}~\bibnamefont {Wu}}, \bibinfo {author}
  {\bibfnamefont {D.~A.}\ \bibnamefont {Muller}}, \bibinfo {author}
  {\bibfnamefont {R.~J.}\ \bibnamefont {Birgeneau}},\ and\ \bibinfo {author}
  {\bibfnamefont {R.}~\bibnamefont {Ramesh}},\ }\bibfield  {title} {\bibinfo
  {title} {Room-temperature skyrmion lattice in a layered magnet
  (fe0.5co0.5)5gete2},\ }\href {https://doi.org/10.1126/sciadv.abm7103}
  {\bibfield  {journal} {\bibinfo  {journal} {Sci. Adv.}\ }\textbf {\bibinfo
  {volume} {8}},\ \bibinfo {pages} {eabm7103} (\bibinfo {year}
  {2022})}\BibitemShut {NoStop}%
\bibitem [{\citenamefont {Peng}\ \emph {et~al.}(2018)\citenamefont {Peng},
  \citenamefont {Zhang}, \citenamefont {Ke}, \citenamefont {Kim}, \citenamefont
  {Zheng}, \citenamefont {Yan}, \citenamefont {Zhang}, \citenamefont {Gao},
  \citenamefont {Wang}, \citenamefont {Cai}, \citenamefont {Shen},
  \citenamefont {McQueeney}, \citenamefont {Kaminski}, \citenamefont {Kramer},\
  and\ \citenamefont {Zhou}}]{Peng2018}%
  \BibitemOpen
  \bibfield  {author} {\bibinfo {author} {\bibfnamefont {L.}~\bibnamefont
  {Peng}}, \bibinfo {author} {\bibfnamefont {Y.}~\bibnamefont {Zhang}},
  \bibinfo {author} {\bibfnamefont {L.}~\bibnamefont {Ke}}, \bibinfo {author}
  {\bibfnamefont {T.-H.}\ \bibnamefont {Kim}}, \bibinfo {author} {\bibfnamefont
  {Q.}~\bibnamefont {Zheng}}, \bibinfo {author} {\bibfnamefont
  {J.}~\bibnamefont {Yan}}, \bibinfo {author} {\bibfnamefont {X.-G.}\
  \bibnamefont {Zhang}}, \bibinfo {author} {\bibfnamefont {Y.}~\bibnamefont
  {Gao}}, \bibinfo {author} {\bibfnamefont {S.}~\bibnamefont {Wang}}, \bibinfo
  {author} {\bibfnamefont {J.}~\bibnamefont {Cai}}, \bibinfo {author}
  {\bibfnamefont {B.}~\bibnamefont {Shen}}, \bibinfo {author} {\bibfnamefont
  {R.~J.}\ \bibnamefont {McQueeney}}, \bibinfo {author} {\bibfnamefont
  {A.}~\bibnamefont {Kaminski}}, \bibinfo {author} {\bibfnamefont {M.~J.}\
  \bibnamefont {Kramer}},\ and\ \bibinfo {author} {\bibfnamefont
  {L.}~\bibnamefont {Zhou}},\ }\bibfield  {title} {\bibinfo {title} {Relaxation
  dynamics of zero-field skyrmions over a wide temperature range},\ }\href
  {https://doi.org/10.1021/acs.nanolett.8b03553} {\bibfield  {journal}
  {\bibinfo  {journal} {Nano Lett.}\ }\textbf {\bibinfo {volume} {18}},\
  \bibinfo {pages} {7777} (\bibinfo {year} {2018})}\BibitemShut {NoStop}%
\bibitem [{\citenamefont {Shinjo}\ \emph {et~al.}(2000)\citenamefont {Shinjo},
  \citenamefont {Okuno}, \citenamefont {Hassdorf}, \citenamefont {Shigeto},\
  and\ \citenamefont {Ono}}]{doi:10.1126/science.289.5481.930}%
  \BibitemOpen
  \bibfield  {author} {\bibinfo {author} {\bibfnamefont {T.}~\bibnamefont
  {Shinjo}}, \bibinfo {author} {\bibfnamefont {T.}~\bibnamefont {Okuno}},
  \bibinfo {author} {\bibfnamefont {R.}~\bibnamefont {Hassdorf}}, \bibinfo
  {author} {\bibfnamefont {K.}~\bibnamefont {Shigeto}},\ and\ \bibinfo {author}
  {\bibfnamefont {T.}~\bibnamefont {Ono}},\ }\bibfield  {title} {\bibinfo
  {title} {Magnetic vortex core observation in circular dots of permalloy},\
  }\href {https://doi.org/10.1126/science.289.5481.930} {\bibfield  {journal}
  {\bibinfo  {journal} {Science}\ }\textbf {\bibinfo {volume} {289}},\ \bibinfo
  {pages} {930} (\bibinfo {year} {2000})}\BibitemShut {NoStop}%
\bibitem [{\citenamefont {Phatak}\ \emph {et~al.}(2012)\citenamefont {Phatak},
  \citenamefont {Petford-Long},\ and\ \citenamefont
  {Heinonen}}]{PhysRevLett.108.067205}%
  \BibitemOpen
  \bibfield  {author} {\bibinfo {author} {\bibfnamefont {C.}~\bibnamefont
  {Phatak}}, \bibinfo {author} {\bibfnamefont {A.~K.}\ \bibnamefont
  {Petford-Long}},\ and\ \bibinfo {author} {\bibfnamefont {O.}~\bibnamefont
  {Heinonen}},\ }\bibfield  {title} {\bibinfo {title} {Direct observation of
  unconventional topological spin structure in coupled magnetic discs},\ }\href
  {https://doi.org/10.1103/PhysRevLett.108.067205} {\bibfield  {journal}
  {\bibinfo  {journal} {Phys. Rev. Lett.}\ }\textbf {\bibinfo {volume} {108}},\
  \bibinfo {pages} {067205} (\bibinfo {year} {2012})}\BibitemShut {NoStop}%
\bibitem [{\citenamefont {Wintz}\ \emph {et~al.}(2013)\citenamefont {Wintz},
  \citenamefont {Bunce}, \citenamefont {Neudert}, \citenamefont {K\"orner},
  \citenamefont {Strache}, \citenamefont {Buhl}, \citenamefont {Erbe},
  \citenamefont {Gemming}, \citenamefont {Raabe}, \citenamefont {Quitmann},\
  and\ \citenamefont {Fassbender}}]{PhysRevLett.110.177201}%
  \BibitemOpen
  \bibfield  {author} {\bibinfo {author} {\bibfnamefont {S.}~\bibnamefont
  {Wintz}}, \bibinfo {author} {\bibfnamefont {C.}~\bibnamefont {Bunce}},
  \bibinfo {author} {\bibfnamefont {A.}~\bibnamefont {Neudert}}, \bibinfo
  {author} {\bibfnamefont {M.}~\bibnamefont {K\"orner}}, \bibinfo {author}
  {\bibfnamefont {T.}~\bibnamefont {Strache}}, \bibinfo {author} {\bibfnamefont
  {M.}~\bibnamefont {Buhl}}, \bibinfo {author} {\bibfnamefont {A.}~\bibnamefont
  {Erbe}}, \bibinfo {author} {\bibfnamefont {S.}~\bibnamefont {Gemming}},
  \bibinfo {author} {\bibfnamefont {J.}~\bibnamefont {Raabe}}, \bibinfo
  {author} {\bibfnamefont {C.}~\bibnamefont {Quitmann}},\ and\ \bibinfo
  {author} {\bibfnamefont {J.}~\bibnamefont {Fassbender}},\ }\bibfield  {title}
  {\bibinfo {title} {Topology and origin of effective spin meron pairs in
  ferromagnetic multilayer elements},\ }\href
  {https://doi.org/10.1103/PhysRevLett.110.177201} {\bibfield  {journal}
  {\bibinfo  {journal} {Phys. Rev. Lett.}\ }\textbf {\bibinfo {volume} {110}},\
  \bibinfo {pages} {177201} (\bibinfo {year} {2013})}\BibitemShut {NoStop}%
\bibitem [{\citenamefont {Tan}\ \emph {et~al.}(2016)\citenamefont {Tan},
  \citenamefont {Li}, \citenamefont {Scholl}, \citenamefont {Arenholz},
  \citenamefont {Young}, \citenamefont {Li}, \citenamefont {Hwang},\ and\
  \citenamefont {Qiu}}]{PhysRevB.94.014433}%
  \BibitemOpen
  \bibfield  {author} {\bibinfo {author} {\bibfnamefont {A.}~\bibnamefont
  {Tan}}, \bibinfo {author} {\bibfnamefont {J.}~\bibnamefont {Li}}, \bibinfo
  {author} {\bibfnamefont {A.}~\bibnamefont {Scholl}}, \bibinfo {author}
  {\bibfnamefont {E.}~\bibnamefont {Arenholz}}, \bibinfo {author}
  {\bibfnamefont {A.~T.}\ \bibnamefont {Young}}, \bibinfo {author}
  {\bibfnamefont {Q.}~\bibnamefont {Li}}, \bibinfo {author} {\bibfnamefont
  {C.}~\bibnamefont {Hwang}},\ and\ \bibinfo {author} {\bibfnamefont {Z.~Q.}\
  \bibnamefont {Qiu}},\ }\bibfield  {title} {\bibinfo {title} {Topology of spin
  meron pairs in coupled {N}i/{F}e/{C}o/{C}u(001) disks},\ }\href
  {https://doi.org/10.1103/PhysRevB.94.014433} {\bibfield  {journal} {\bibinfo
  {journal} {Phys. Rev. B}\ }\textbf {\bibinfo {volume} {94}},\ \bibinfo
  {pages} {014433} (\bibinfo {year} {2016})}\BibitemShut {NoStop}%
\bibitem [{\citenamefont {Shigeto}\ \emph {et~al.}(2002)\citenamefont
  {Shigeto}, \citenamefont {Okuno}, \citenamefont {Mibu}, \citenamefont
  {Shinjo},\ and\ \citenamefont {Ono}}]{10.1063/1.1483386}%
  \BibitemOpen
  \bibfield  {author} {\bibinfo {author} {\bibfnamefont {K.}~\bibnamefont
  {Shigeto}}, \bibinfo {author} {\bibfnamefont {T.}~\bibnamefont {Okuno}},
  \bibinfo {author} {\bibfnamefont {K.}~\bibnamefont {Mibu}}, \bibinfo {author}
  {\bibfnamefont {T.}~\bibnamefont {Shinjo}},\ and\ \bibinfo {author}
  {\bibfnamefont {T.}~\bibnamefont {Ono}},\ }\bibfield  {title} {\bibinfo
  {title} {Magnetic force microscopy observation of antivortex core with
  perpendicular magnetization in patterned thin film of permalloy},\ }\href
  {https://doi.org/10.1063/1.1483386} {\bibfield  {journal} {\bibinfo
  {journal} {App. Phys. Lett.}\ }\textbf {\bibinfo {volume} {80}},\ \bibinfo
  {pages} {4190} (\bibinfo {year} {2002})}\BibitemShut {NoStop}%
\bibitem [{\citenamefont {Fu}\ \emph {et~al.}(2018)\citenamefont {Fu},
  \citenamefont {Pollard}, \citenamefont {Chen}, \citenamefont {Yoo},
  \citenamefont {Yang},\ and\ \citenamefont
  {Zhu}}]{doi:10.1126/sciadv.aat3077}%
  \BibitemOpen
  \bibfield  {author} {\bibinfo {author} {\bibfnamefont {X.}~\bibnamefont
  {Fu}}, \bibinfo {author} {\bibfnamefont {S.~D.}\ \bibnamefont {Pollard}},
  \bibinfo {author} {\bibfnamefont {B.}~\bibnamefont {Chen}}, \bibinfo {author}
  {\bibfnamefont {B.-K.}\ \bibnamefont {Yoo}}, \bibinfo {author} {\bibfnamefont
  {H.}~\bibnamefont {Yang}},\ and\ \bibinfo {author} {\bibfnamefont
  {Y.}~\bibnamefont {Zhu}},\ }\bibfield  {title} {\bibinfo {title} {Optical
  manipulation of magnetic vortices visualized in situ by lorentz electron
  microscopy},\ }\href {https://doi.org/10.1126/sciadv.aat3077} {\bibfield
  {journal} {\bibinfo  {journal} {Sci. Adv.}\ }\textbf {\bibinfo {volume}
  {4}},\ \bibinfo {pages} {eaat3077} (\bibinfo {year} {2018})}\BibitemShut
  {NoStop}%
\bibitem [{\citenamefont {Van~Waeyenberge}\ \emph {et~al.}(2006)\citenamefont
  {Van~Waeyenberge}, \citenamefont {Puzic}, \citenamefont {Stoll},
  \citenamefont {Chou}, \citenamefont {Tyliszczak}, \citenamefont {Hertel},
  \citenamefont {F{\"a}hnle}, \citenamefont {Br{\"u}ckl}, \citenamefont {Rott},
  \citenamefont {Reiss}, \citenamefont {Neudecker}, \citenamefont {Weiss},
  \citenamefont {Back},\ and\ \citenamefont {Sch{\"u}tz}}]{VanWaeyenberge2006}%
  \BibitemOpen
  \bibfield  {author} {\bibinfo {author} {\bibfnamefont {B.}~\bibnamefont
  {Van~Waeyenberge}}, \bibinfo {author} {\bibfnamefont {A.}~\bibnamefont
  {Puzic}}, \bibinfo {author} {\bibfnamefont {H.}~\bibnamefont {Stoll}},
  \bibinfo {author} {\bibfnamefont {K.~W.}\ \bibnamefont {Chou}}, \bibinfo
  {author} {\bibfnamefont {T.}~\bibnamefont {Tyliszczak}}, \bibinfo {author}
  {\bibfnamefont {R.}~\bibnamefont {Hertel}}, \bibinfo {author} {\bibfnamefont
  {M.}~\bibnamefont {F{\"a}hnle}}, \bibinfo {author} {\bibfnamefont
  {H.}~\bibnamefont {Br{\"u}ckl}}, \bibinfo {author} {\bibfnamefont
  {K.}~\bibnamefont {Rott}}, \bibinfo {author} {\bibfnamefont {G.}~\bibnamefont
  {Reiss}}, \bibinfo {author} {\bibfnamefont {I.}~\bibnamefont {Neudecker}},
  \bibinfo {author} {\bibfnamefont {D.}~\bibnamefont {Weiss}}, \bibinfo
  {author} {\bibfnamefont {C.~H.}\ \bibnamefont {Back}},\ and\ \bibinfo
  {author} {\bibfnamefont {G.}~\bibnamefont {Sch{\"u}tz}},\ }\bibfield  {title}
  {\bibinfo {title} {Magnetic vortex core reversal by excitation with short
  bursts of an alternating field},\ }\href
  {https://doi.org/10.1038/nature05240} {\bibfield  {journal} {\bibinfo
  {journal} {Nature}\ }\textbf {\bibinfo {volume} {444}},\ \bibinfo {pages}
  {461} (\bibinfo {year} {2006})}\BibitemShut {NoStop}%
\bibitem [{\citenamefont {Ruotolo}\ \emph {et~al.}(2009)\citenamefont
  {Ruotolo}, \citenamefont {Cros}, \citenamefont {Georges}, \citenamefont
  {Dussaux}, \citenamefont {Grollier}, \citenamefont {Deranlot}, \citenamefont
  {Guillemet}, \citenamefont {Bouzehouane}, \citenamefont {Fusil},\ and\
  \citenamefont {Fert}}]{Ruotolo2009}%
  \BibitemOpen
  \bibfield  {author} {\bibinfo {author} {\bibfnamefont {A.}~\bibnamefont
  {Ruotolo}}, \bibinfo {author} {\bibfnamefont {V.}~\bibnamefont {Cros}},
  \bibinfo {author} {\bibfnamefont {B.}~\bibnamefont {Georges}}, \bibinfo
  {author} {\bibfnamefont {A.}~\bibnamefont {Dussaux}}, \bibinfo {author}
  {\bibfnamefont {J.}~\bibnamefont {Grollier}}, \bibinfo {author}
  {\bibfnamefont {C.}~\bibnamefont {Deranlot}}, \bibinfo {author}
  {\bibfnamefont {R.}~\bibnamefont {Guillemet}}, \bibinfo {author}
  {\bibfnamefont {K.}~\bibnamefont {Bouzehouane}}, \bibinfo {author}
  {\bibfnamefont {S.}~\bibnamefont {Fusil}},\ and\ \bibinfo {author}
  {\bibfnamefont {A.}~\bibnamefont {Fert}},\ }\bibfield  {title} {\bibinfo
  {title} {Phase-locking of magnetic vortices mediated by antivortices},\
  }\href {https://doi.org/10.1038/nnano.2009.143} {\bibfield  {journal}
  {\bibinfo  {journal} {Nat. Nanotechnol.}\ }\textbf {\bibinfo {volume} {4}},\
  \bibinfo {pages} {528} (\bibinfo {year} {2009})}\BibitemShut {NoStop}%
\bibitem [{\citenamefont {Roy}\ \emph {et~al.}(2009)\citenamefont {Roy},
  \citenamefont {Lee}, \citenamefont {Trypiniotis}, \citenamefont {Anderson},
  \citenamefont {Jones}, \citenamefont {Tse},\ and\ \citenamefont
  {Barnes}}]{PhysRevB.79.060407}%
  \BibitemOpen
  \bibfield  {author} {\bibinfo {author} {\bibfnamefont {P.~E.}\ \bibnamefont
  {Roy}}, \bibinfo {author} {\bibfnamefont {J.~H.}\ \bibnamefont {Lee}},
  \bibinfo {author} {\bibfnamefont {T.}~\bibnamefont {Trypiniotis}}, \bibinfo
  {author} {\bibfnamefont {D.}~\bibnamefont {Anderson}}, \bibinfo {author}
  {\bibfnamefont {G.~A.~C.}\ \bibnamefont {Jones}}, \bibinfo {author}
  {\bibfnamefont {D.}~\bibnamefont {Tse}},\ and\ \bibinfo {author}
  {\bibfnamefont {C.~H.~W.}\ \bibnamefont {Barnes}},\ }\bibfield  {title}
  {\bibinfo {title} {Antivortex domain walls observed in permalloy rings via
  magnetic force microscopy},\ }\href
  {https://doi.org/10.1103/PhysRevB.79.060407} {\bibfield  {journal} {\bibinfo
  {journal} {Phys. Rev. B}\ }\textbf {\bibinfo {volume} {79}},\ \bibinfo
  {pages} {060407} (\bibinfo {year} {2009})}\BibitemShut {NoStop}%
\bibitem [{\citenamefont {Chmiel}\ \emph {et~al.}(2018)\citenamefont {Chmiel},
  \citenamefont {Waterfield~Price}, \citenamefont {Johnson}, \citenamefont
  {Lamirand}, \citenamefont {Schad}, \citenamefont {van~der Laan},
  \citenamefont {Harris}, \citenamefont {Irwin}, \citenamefont {Rzchowski},
  \citenamefont {Eom},\ and\ \citenamefont {Radaelli}}]{Chmiel2018}%
  \BibitemOpen
  \bibfield  {author} {\bibinfo {author} {\bibfnamefont {F.~P.}\ \bibnamefont
  {Chmiel}}, \bibinfo {author} {\bibfnamefont {N.}~\bibnamefont
  {Waterfield~Price}}, \bibinfo {author} {\bibfnamefont {R.~D.}\ \bibnamefont
  {Johnson}}, \bibinfo {author} {\bibfnamefont {A.~D.}\ \bibnamefont
  {Lamirand}}, \bibinfo {author} {\bibfnamefont {J.}~\bibnamefont {Schad}},
  \bibinfo {author} {\bibfnamefont {G.}~\bibnamefont {van~der Laan}}, \bibinfo
  {author} {\bibfnamefont {D.~T.}\ \bibnamefont {Harris}}, \bibinfo {author}
  {\bibfnamefont {J.}~\bibnamefont {Irwin}}, \bibinfo {author} {\bibfnamefont
  {M.~S.}\ \bibnamefont {Rzchowski}}, \bibinfo {author} {\bibfnamefont {C.-B.}\
  \bibnamefont {Eom}},\ and\ \bibinfo {author} {\bibfnamefont {P.~G.}\
  \bibnamefont {Radaelli}},\ }\bibfield  {title} {\bibinfo {title} {Observation
  of magnetic vortex pairs at room temperature in a planar
  $\alpha$-{F}e\textsubscript{2}{O}\textsubscript{3}/{C}o heterostructure},\
  }\href {https://doi.org/10.1038/s41563-018-0101-x} {\bibfield  {journal}
  {\bibinfo  {journal} {Nat. Mater.}\ }\textbf {\bibinfo {volume} {17}},\
  \bibinfo {pages} {581} (\bibinfo {year} {2018})}\BibitemShut {NoStop}%
\bibitem [{\citenamefont {Gao}\ \emph {et~al.}(2019)\citenamefont {Gao},
  \citenamefont {Je}, \citenamefont {Im}, \citenamefont {Choi}, \citenamefont
  {Yang}, \citenamefont {Li}, \citenamefont {Wang}, \citenamefont {Lee},
  \citenamefont {Han}, \citenamefont {Lee}, \citenamefont {Chao}, \citenamefont
  {Hwang}, \citenamefont {Li},\ and\ \citenamefont {Qiu}}]{Gao2019}%
  \BibitemOpen
  \bibfield  {author} {\bibinfo {author} {\bibfnamefont {N.}~\bibnamefont
  {Gao}}, \bibinfo {author} {\bibfnamefont {S.-G.}\ \bibnamefont {Je}},
  \bibinfo {author} {\bibfnamefont {M.-Y.}\ \bibnamefont {Im}}, \bibinfo
  {author} {\bibfnamefont {J.~W.}\ \bibnamefont {Choi}}, \bibinfo {author}
  {\bibfnamefont {M.}~\bibnamefont {Yang}}, \bibinfo {author} {\bibfnamefont
  {Q.}~\bibnamefont {Li}}, \bibinfo {author} {\bibfnamefont {T.~Y.}\
  \bibnamefont {Wang}}, \bibinfo {author} {\bibfnamefont {S.}~\bibnamefont
  {Lee}}, \bibinfo {author} {\bibfnamefont {H.-S.}\ \bibnamefont {Han}},
  \bibinfo {author} {\bibfnamefont {K.-S.}\ \bibnamefont {Lee}}, \bibinfo
  {author} {\bibfnamefont {W.}~\bibnamefont {Chao}}, \bibinfo {author}
  {\bibfnamefont {C.}~\bibnamefont {Hwang}}, \bibinfo {author} {\bibfnamefont
  {J.}~\bibnamefont {Li}},\ and\ \bibinfo {author} {\bibfnamefont {Z.~Q.}\
  \bibnamefont {Qiu}},\ }\bibfield  {title} {\bibinfo {title} {Creation and
  annihilation of topological meron pairs in in-plane magnetized films},\
  }\href {https://doi.org/10.1038/s41467-019-13642-z} {\bibfield  {journal}
  {\bibinfo  {journal} {Nat. Commun.}\ }\textbf {\bibinfo {volume} {10}},\
  \bibinfo {pages} {5603} (\bibinfo {year} {2019})}\BibitemShut {NoStop}%
\bibitem [{\citenamefont {Lu}\ \emph {et~al.}(2020)\citenamefont {Lu},
  \citenamefont {Fei}, \citenamefont {Zhu},\ and\ \citenamefont
  {Yang}}]{Lu2020}%
  \BibitemOpen
  \bibfield  {author} {\bibinfo {author} {\bibfnamefont {X.}~\bibnamefont
  {Lu}}, \bibinfo {author} {\bibfnamefont {R.}~\bibnamefont {Fei}}, \bibinfo
  {author} {\bibfnamefont {L.}~\bibnamefont {Zhu}},\ and\ \bibinfo {author}
  {\bibfnamefont {L.}~\bibnamefont {Yang}},\ }\bibfield  {title} {\bibinfo
  {title} {Meron-like topological spin defects in monolayer
  {C}r{C}l\textsubscript{3}},\ }\href
  {https://doi.org/10.1038/s41467-020-18573-8} {\bibfield  {journal} {\bibinfo
  {journal} {Nat. Commun.}\ }\textbf {\bibinfo {volume} {11}},\ \bibinfo
  {pages} {4724} (\bibinfo {year} {2020})}\BibitemShut {NoStop}%
\bibitem [{\citenamefont {Augustin}\ \emph {et~al.}(2021)\citenamefont
  {Augustin}, \citenamefont {Jenkins}, \citenamefont {Evans}, \citenamefont
  {Novoselov},\ and\ \citenamefont {Santos}}]{Augustin2021}%
  \BibitemOpen
  \bibfield  {author} {\bibinfo {author} {\bibfnamefont {M.}~\bibnamefont
  {Augustin}}, \bibinfo {author} {\bibfnamefont {S.}~\bibnamefont {Jenkins}},
  \bibinfo {author} {\bibfnamefont {R.~F.~L.}\ \bibnamefont {Evans}}, \bibinfo
  {author} {\bibfnamefont {K.~S.}\ \bibnamefont {Novoselov}},\ and\ \bibinfo
  {author} {\bibfnamefont {E.~J.~G.}\ \bibnamefont {Santos}},\ }\bibfield
  {title} {\bibinfo {title} {Properties and dynamics of meron topological spin
  textures in the two-dimensional magnet {C}r{C}l\textsubscript{3}},\ }\href
  {https://doi.org/10.1038/s41467-020-20497-2} {\bibfield  {journal} {\bibinfo
  {journal} {Nat. Commun.}\ }\textbf {\bibinfo {volume} {12}},\ \bibinfo
  {pages} {185} (\bibinfo {year} {2021})}\BibitemShut {NoStop}%
\bibitem [{\citenamefont {Tong}\ \emph {et~al.}(2018)\citenamefont {Tong},
  \citenamefont {Liu}, \citenamefont {Xiao},\ and\ \citenamefont
  {Yao}}]{Tong2018}%
  \BibitemOpen
  \bibfield  {author} {\bibinfo {author} {\bibfnamefont {Q.}~\bibnamefont
  {Tong}}, \bibinfo {author} {\bibfnamefont {F.}~\bibnamefont {Liu}}, \bibinfo
  {author} {\bibfnamefont {J.}~\bibnamefont {Xiao}},\ and\ \bibinfo {author}
  {\bibfnamefont {W.}~\bibnamefont {Yao}},\ }\bibfield  {title} {\bibinfo
  {title} {Skyrmions in the moiré of van der {W}aals 2{D} magnets},\ }\href
  {https://doi.org/10.1021/acs.nanolett.8b03315} {\bibfield  {journal}
  {\bibinfo  {journal} {Nano Lett.}\ }\textbf {\bibinfo {volume} {18}},\
  \bibinfo {pages} {7194} (\bibinfo {year} {2018})}\BibitemShut {NoStop}%
\bibitem [{\citenamefont {Akram}\ and\ \citenamefont
  {Erten}(2021)}]{PhysRevB.103.L140406}%
  \BibitemOpen
  \bibfield  {author} {\bibinfo {author} {\bibfnamefont {M.}~\bibnamefont
  {Akram}}\ and\ \bibinfo {author} {\bibfnamefont {O.}~\bibnamefont {Erten}},\
  }\bibfield  {title} {\bibinfo {title} {Skyrmions in twisted van der {W}aals
  magnets},\ }\href {https://doi.org/10.1103/PhysRevB.103.L140406} {\bibfield
  {journal} {\bibinfo  {journal} {Phys. Rev. B}\ }\textbf {\bibinfo {volume}
  {103}},\ \bibinfo {pages} {L140406} (\bibinfo {year} {2021})}\BibitemShut
  {NoStop}%
\bibitem [{\citenamefont {Ghader}\ \emph {et~al.}(2022)\citenamefont {Ghader},
  \citenamefont {Jabakhanji},\ and\ \citenamefont {Stroppa}}]{Ghader2022}%
  \BibitemOpen
  \bibfield  {author} {\bibinfo {author} {\bibfnamefont {D.}~\bibnamefont
  {Ghader}}, \bibinfo {author} {\bibfnamefont {B.}~\bibnamefont {Jabakhanji}},\
  and\ \bibinfo {author} {\bibfnamefont {A.}~\bibnamefont {Stroppa}},\
  }\bibfield  {title} {\bibinfo {title} {Whirling interlayer fields as a source
  of stable topological order in moir{\'e} {C}r{I}\textsubscript{3}},\ }\href
  {https://doi.org/10.1038/s42005-022-00972-6} {\bibfield  {journal} {\bibinfo
  {journal} {Commun. Phys.}\ }\textbf {\bibinfo {volume} {5}},\ \bibinfo
  {pages} {192} (\bibinfo {year} {2022})}\BibitemShut {NoStop}%
\bibitem [{\citenamefont {Kim}\ \emph {et~al.}(2023)\citenamefont {Kim},
  \citenamefont {Kiem}, \citenamefont {Bednik}, \citenamefont {Han},\ and\
  \citenamefont {Park}}]{Kim2023}%
  \BibitemOpen
  \bibfield  {author} {\bibinfo {author} {\bibfnamefont {K.-M.}\ \bibnamefont
  {Kim}}, \bibinfo {author} {\bibfnamefont {D.~H.}\ \bibnamefont {Kiem}},
  \bibinfo {author} {\bibfnamefont {G.}~\bibnamefont {Bednik}}, \bibinfo
  {author} {\bibfnamefont {M.~J.}\ \bibnamefont {Han}},\ and\ \bibinfo {author}
  {\bibfnamefont {M.~J.}\ \bibnamefont {Park}},\ }\bibfield  {title} {\bibinfo
  {title} {Ab initio spin hamiltonian and topological noncentrosymmetric
  magnetism in twisted bilayer {C}r{I}3},\ }\href
  {https://doi.org/10.1021/acs.nanolett.3c01529} {\bibfield  {journal}
  {\bibinfo  {journal} {Nano Lett.}\ }\textbf {\bibinfo {volume} {23}},\
  \bibinfo {pages} {6088} (\bibinfo {year} {2023})}\BibitemShut {NoStop}%
\bibitem [{\citenamefont {Kim}\ and\ \citenamefont
  {Park}(2023)}]{PhysRevB.108.L100401}%
  \BibitemOpen
  \bibfield  {author} {\bibinfo {author} {\bibfnamefont {K.-M.}\ \bibnamefont
  {Kim}}\ and\ \bibinfo {author} {\bibfnamefont {M.~J.}\ \bibnamefont {Park}},\
  }\bibfield  {title} {\bibinfo {title} {Controllable magnetic domains in
  twisted trilayer magnets},\ }\href
  {https://doi.org/10.1103/PhysRevB.108.L100401} {\bibfield  {journal}
  {\bibinfo  {journal} {Phys. Rev. B}\ }\textbf {\bibinfo {volume} {108}},\
  \bibinfo {pages} {L100401} (\bibinfo {year} {2023})}\BibitemShut {NoStop}%
\bibitem [{\citenamefont {Wang}\ \emph {et~al.}(2022)\citenamefont {Wang},
  \citenamefont {Bedoya-Pinto}, \citenamefont {Blei}, \citenamefont {Dismukes},
  \citenamefont {Hamo}, \citenamefont {Jenkins}, \citenamefont {Koperski},
  \citenamefont {Liu}, \citenamefont {Sun}, \citenamefont {Telford},
  \citenamefont {Kim}, \citenamefont {Augustin}, \citenamefont {Vool},
  \citenamefont {Yin}, \citenamefont {Li}, \citenamefont {Falin}, \citenamefont
  {Dean}, \citenamefont {Casanova}, \citenamefont {Evans}, \citenamefont
  {Chshiev}, \citenamefont {Mishchenko}, \citenamefont {Petrovic},
  \citenamefont {He}, \citenamefont {Zhao}, \citenamefont {Tsen}, \citenamefont
  {Gerardot}, \citenamefont {Brotons-Gisbert}, \citenamefont {Guguchia},
  \citenamefont {Roy}, \citenamefont {Tongay}, \citenamefont {Wang},
  \citenamefont {Hasan}, \citenamefont {Wrachtrup}, \citenamefont {Yacoby},
  \citenamefont {Fert}, \citenamefont {Parkin}, \citenamefont {Novoselov},
  \citenamefont {Dai}, \citenamefont {Balicas},\ and\ \citenamefont
  {Santos}}]{Wang2022}%
  \BibitemOpen
  \bibfield  {author} {\bibinfo {author} {\bibfnamefont {Q.~H.}\ \bibnamefont
  {Wang}}, \bibinfo {author} {\bibfnamefont {A.}~\bibnamefont {Bedoya-Pinto}},
  \bibinfo {author} {\bibfnamefont {M.}~\bibnamefont {Blei}}, \bibinfo {author}
  {\bibfnamefont {A.~H.}\ \bibnamefont {Dismukes}}, \bibinfo {author}
  {\bibfnamefont {A.}~\bibnamefont {Hamo}}, \bibinfo {author} {\bibfnamefont
  {S.}~\bibnamefont {Jenkins}}, \bibinfo {author} {\bibfnamefont
  {M.}~\bibnamefont {Koperski}}, \bibinfo {author} {\bibfnamefont
  {Y.}~\bibnamefont {Liu}}, \bibinfo {author} {\bibfnamefont {Q.-C.}\
  \bibnamefont {Sun}}, \bibinfo {author} {\bibfnamefont {E.~J.}\ \bibnamefont
  {Telford}}, \bibinfo {author} {\bibfnamefont {H.~H.}\ \bibnamefont {Kim}},
  \bibinfo {author} {\bibfnamefont {M.}~\bibnamefont {Augustin}}, \bibinfo
  {author} {\bibfnamefont {U.}~\bibnamefont {Vool}}, \bibinfo {author}
  {\bibfnamefont {J.-X.}\ \bibnamefont {Yin}}, \bibinfo {author} {\bibfnamefont
  {L.~H.}\ \bibnamefont {Li}}, \bibinfo {author} {\bibfnamefont
  {A.}~\bibnamefont {Falin}}, \bibinfo {author} {\bibfnamefont {C.~R.}\
  \bibnamefont {Dean}}, \bibinfo {author} {\bibfnamefont {F.}~\bibnamefont
  {Casanova}}, \bibinfo {author} {\bibfnamefont {R.~F.~L.}\ \bibnamefont
  {Evans}}, \bibinfo {author} {\bibfnamefont {M.}~\bibnamefont {Chshiev}},
  \bibinfo {author} {\bibfnamefont {A.}~\bibnamefont {Mishchenko}}, \bibinfo
  {author} {\bibfnamefont {C.}~\bibnamefont {Petrovic}}, \bibinfo {author}
  {\bibfnamefont {R.}~\bibnamefont {He}}, \bibinfo {author} {\bibfnamefont
  {L.}~\bibnamefont {Zhao}}, \bibinfo {author} {\bibfnamefont {A.~W.}\
  \bibnamefont {Tsen}}, \bibinfo {author} {\bibfnamefont {B.~D.}\ \bibnamefont
  {Gerardot}}, \bibinfo {author} {\bibfnamefont {M.}~\bibnamefont
  {Brotons-Gisbert}}, \bibinfo {author} {\bibfnamefont {Z.}~\bibnamefont
  {Guguchia}}, \bibinfo {author} {\bibfnamefont {X.}~\bibnamefont {Roy}},
  \bibinfo {author} {\bibfnamefont {S.}~\bibnamefont {Tongay}}, \bibinfo
  {author} {\bibfnamefont {Z.}~\bibnamefont {Wang}}, \bibinfo {author}
  {\bibfnamefont {M.~Z.}\ \bibnamefont {Hasan}}, \bibinfo {author}
  {\bibfnamefont {J.}~\bibnamefont {Wrachtrup}}, \bibinfo {author}
  {\bibfnamefont {A.}~\bibnamefont {Yacoby}}, \bibinfo {author} {\bibfnamefont
  {A.}~\bibnamefont {Fert}}, \bibinfo {author} {\bibfnamefont {S.}~\bibnamefont
  {Parkin}}, \bibinfo {author} {\bibfnamefont {K.~S.}\ \bibnamefont
  {Novoselov}}, \bibinfo {author} {\bibfnamefont {P.}~\bibnamefont {Dai}},
  \bibinfo {author} {\bibfnamefont {L.}~\bibnamefont {Balicas}},\ and\ \bibinfo
  {author} {\bibfnamefont {E.~J.~G.}\ \bibnamefont {Santos}},\ }\bibfield
  {title} {\bibinfo {title} {The magnetic genome of two-dimensional van der
  waals materials},\ }\href {https://doi.org/10.1021/acsnano.1c09150}
  {\bibfield  {journal} {\bibinfo  {journal} {ACS Nano}\ }\textbf {\bibinfo
  {volume} {16}},\ \bibinfo {pages} {6960} (\bibinfo {year}
  {2022})}\BibitemShut {NoStop}%
\bibitem [{SI()}]{SI}%
  \BibitemOpen
  \href@noop {} {}\bibinfo {note} {See Supporting Information}\BibitemShut
  {NoStop}%
\bibitem [{\citenamefont {Zou}\ \emph {et~al.}(2019)\citenamefont {Zou},
  \citenamefont {Kim},\ and\ \citenamefont {Tserkovnyak}}]{PhysRevB.99.180402}%
  \BibitemOpen
  \bibfield  {author} {\bibinfo {author} {\bibfnamefont {J.}~\bibnamefont
  {Zou}}, \bibinfo {author} {\bibfnamefont {S.~K.}\ \bibnamefont {Kim}},\ and\
  \bibinfo {author} {\bibfnamefont {Y.}~\bibnamefont {Tserkovnyak}},\
  }\bibfield  {title} {\bibinfo {title} {Topological transport of vorticity in
  heisenberg magnets},\ }\href {https://doi.org/10.1103/PhysRevB.99.180402}
  {\bibfield  {journal} {\bibinfo  {journal} {Phys. Rev. B}\ }\textbf {\bibinfo
  {volume} {99}},\ \bibinfo {pages} {180402} (\bibinfo {year}
  {2019})}\BibitemShut {NoStop}%
\bibitem [{\citenamefont {Hubert}\ and\ \citenamefont
  {Schäfer}(1998)}]{Hubert1998}%
  \BibitemOpen
  \bibfield  {author} {\bibinfo {author} {\bibfnamefont {A.}~\bibnamefont
  {Hubert}}\ and\ \bibinfo {author} {\bibfnamefont {R.}~\bibnamefont
  {Schäfer}},\ }\href {https://doi.org/10.1007/978-3-540-85054-0} {\emph
  {\bibinfo {title} {Magnetic Domains}}}\ (\bibinfo  {publisher} {Springer
  Berlin Heidelberg},\ \bibinfo {year} {1998})\BibitemShut {NoStop}%
\bibitem [{\citenamefont {Song}\ \emph {et~al.}(2021)\citenamefont {Song},
  \citenamefont {Sun}, \citenamefont {Anderson}, \citenamefont {Wang},
  \citenamefont {Qian}, \citenamefont {Taniguchi}, \citenamefont {Watanabe},
  \citenamefont {McGuire}, \citenamefont {Stöhr}, \citenamefont {Xiao},
  \citenamefont {Cao}, \citenamefont {Wrachtrup},\ and\ \citenamefont
  {Xu}}]{Song2021}%
  \BibitemOpen
  \bibfield  {author} {\bibinfo {author} {\bibfnamefont {T.}~\bibnamefont
  {Song}}, \bibinfo {author} {\bibfnamefont {Q.-C.}\ \bibnamefont {Sun}},
  \bibinfo {author} {\bibfnamefont {E.}~\bibnamefont {Anderson}}, \bibinfo
  {author} {\bibfnamefont {C.}~\bibnamefont {Wang}}, \bibinfo {author}
  {\bibfnamefont {J.}~\bibnamefont {Qian}}, \bibinfo {author} {\bibfnamefont
  {T.}~\bibnamefont {Taniguchi}}, \bibinfo {author} {\bibfnamefont
  {K.}~\bibnamefont {Watanabe}}, \bibinfo {author} {\bibfnamefont {M.~A.}\
  \bibnamefont {McGuire}}, \bibinfo {author} {\bibfnamefont {R.}~\bibnamefont
  {Stöhr}}, \bibinfo {author} {\bibfnamefont {D.}~\bibnamefont {Xiao}},
  \bibinfo {author} {\bibfnamefont {T.}~\bibnamefont {Cao}}, \bibinfo {author}
  {\bibfnamefont {J.}~\bibnamefont {Wrachtrup}},\ and\ \bibinfo {author}
  {\bibfnamefont {X.}~\bibnamefont {Xu}},\ }\bibfield  {title} {\bibinfo
  {title} {Direct visualization of magnetic domains and moiré magnetism in
  twisted 2{D} magnets},\ }\href {https://doi.org/10.1126/science.abj7478}
  {\bibfield  {journal} {\bibinfo  {journal} {Science}\ }\textbf {\bibinfo
  {volume} {374}},\ \bibinfo {pages} {1140} (\bibinfo {year}
  {2021})}\BibitemShut {NoStop}%
\bibitem [{\citenamefont {Xu}\ \emph {et~al.}(2022)\citenamefont {Xu},
  \citenamefont {Ray}, \citenamefont {Shao}, \citenamefont {Jiang},
  \citenamefont {Lee}, \citenamefont {Weber}, \citenamefont {Goldberger},
  \citenamefont {Watanabe}, \citenamefont {Taniguchi}, \citenamefont {Muller},
  \citenamefont {Mak},\ and\ \citenamefont {Shan}}]{Xu2022}%
  \BibitemOpen
  \bibfield  {author} {\bibinfo {author} {\bibfnamefont {Y.}~\bibnamefont
  {Xu}}, \bibinfo {author} {\bibfnamefont {A.}~\bibnamefont {Ray}}, \bibinfo
  {author} {\bibfnamefont {Y.-T.}\ \bibnamefont {Shao}}, \bibinfo {author}
  {\bibfnamefont {S.}~\bibnamefont {Jiang}}, \bibinfo {author} {\bibfnamefont
  {K.}~\bibnamefont {Lee}}, \bibinfo {author} {\bibfnamefont {D.}~\bibnamefont
  {Weber}}, \bibinfo {author} {\bibfnamefont {J.~E.}\ \bibnamefont
  {Goldberger}}, \bibinfo {author} {\bibfnamefont {K.}~\bibnamefont
  {Watanabe}}, \bibinfo {author} {\bibfnamefont {T.}~\bibnamefont {Taniguchi}},
  \bibinfo {author} {\bibfnamefont {D.~A.}\ \bibnamefont {Muller}}, \bibinfo
  {author} {\bibfnamefont {K.~F.}\ \bibnamefont {Mak}},\ and\ \bibinfo {author}
  {\bibfnamefont {J.}~\bibnamefont {Shan}},\ }\bibfield  {title} {\bibinfo
  {title} {Coexisting ferromagnetic--antiferromagnetic state in twisted bilayer
  {C}r{I}\textsubscript{3}},\ }\href
  {https://doi.org/10.1038/s41565-021-01014-y} {\bibfield  {journal} {\bibinfo
  {journal} {Nat. Nanotechnol.}\ }\textbf {\bibinfo {volume} {17}},\ \bibinfo
  {pages} {143} (\bibinfo {year} {2022})}\BibitemShut {NoStop}%
\bibitem [{\citenamefont {Xie}\ \emph {et~al.}(2023)\citenamefont {Xie},
  \citenamefont {Luo}, \citenamefont {Ye}, \citenamefont {Sun}, \citenamefont
  {Ye}, \citenamefont {Sung}, \citenamefont {Ge}, \citenamefont {Yan},
  \citenamefont {Fu}, \citenamefont {Tian}, \citenamefont {Lei}, \citenamefont
  {Sun}, \citenamefont {Hovden}, \citenamefont {He},\ and\ \citenamefont
  {Zhao}}]{Xie2023}%
  \BibitemOpen
  \bibfield  {author} {\bibinfo {author} {\bibfnamefont {H.}~\bibnamefont
  {Xie}}, \bibinfo {author} {\bibfnamefont {X.}~\bibnamefont {Luo}}, \bibinfo
  {author} {\bibfnamefont {Z.}~\bibnamefont {Ye}}, \bibinfo {author}
  {\bibfnamefont {Z.}~\bibnamefont {Sun}}, \bibinfo {author} {\bibfnamefont
  {G.}~\bibnamefont {Ye}}, \bibinfo {author} {\bibfnamefont {S.~H.}\
  \bibnamefont {Sung}}, \bibinfo {author} {\bibfnamefont {H.}~\bibnamefont
  {Ge}}, \bibinfo {author} {\bibfnamefont {S.}~\bibnamefont {Yan}}, \bibinfo
  {author} {\bibfnamefont {Y.}~\bibnamefont {Fu}}, \bibinfo {author}
  {\bibfnamefont {S.}~\bibnamefont {Tian}}, \bibinfo {author} {\bibfnamefont
  {H.}~\bibnamefont {Lei}}, \bibinfo {author} {\bibfnamefont {K.}~\bibnamefont
  {Sun}}, \bibinfo {author} {\bibfnamefont {R.}~\bibnamefont {Hovden}},
  \bibinfo {author} {\bibfnamefont {R.}~\bibnamefont {He}},\ and\ \bibinfo
  {author} {\bibfnamefont {L.}~\bibnamefont {Zhao}},\ }\bibfield  {title}
  {\bibinfo {title} {Evidence of non-collinear spin texture in magnetic
  moir{\'e} superlattices},\ }\href
  {https://doi.org/10.1038/s41567-023-02061-z} {\bibfield  {journal} {\bibinfo
  {journal} {Nat. Phys.}\ }\textbf {\bibinfo {volume} {19}},\ \bibinfo {pages}
  {1150} (\bibinfo {year} {2023})}\BibitemShut {NoStop}%
\bibitem [{\citenamefont {Jiao}\ \emph {et~al.}(2023)\citenamefont {Jiao},
  \citenamefont {Pei}, \citenamefont {Wu}, \citenamefont {Wang},\ and\
  \citenamefont {Xia}}]{Jiao_2023}%
  \BibitemOpen
  \bibfield  {author} {\bibinfo {author} {\bibfnamefont {C.}~\bibnamefont
  {Jiao}}, \bibinfo {author} {\bibfnamefont {S.}~\bibnamefont {Pei}}, \bibinfo
  {author} {\bibfnamefont {S.}~\bibnamefont {Wu}}, \bibinfo {author}
  {\bibfnamefont {Z.}~\bibnamefont {Wang}},\ and\ \bibinfo {author}
  {\bibfnamefont {J.}~\bibnamefont {Xia}},\ }\bibfield  {title} {\bibinfo
  {title} {Tuning and exploiting interlayer coupling in two-dimensional van der
  waals heterostructures},\ }\href {https://doi.org/10.1088/1361-6633/acfe89}
  {\bibfield  {journal} {\bibinfo  {journal} {Rep. Prog. Phys.}\ }\textbf
  {\bibinfo {volume} {86}},\ \bibinfo {pages} {114503} (\bibinfo {year}
  {2023})}\BibitemShut {NoStop}%
\bibitem [{\citenamefont {Jena}\ \emph {et~al.}(2020)\citenamefont {Jena},
  \citenamefont {G{\"o}bel}, \citenamefont {Ma}, \citenamefont {Kumar},
  \citenamefont {Saha}, \citenamefont {Mertig}, \citenamefont {Felser},\ and\
  \citenamefont {Parkin}}]{Jena2020}%
  \BibitemOpen
  \bibfield  {author} {\bibinfo {author} {\bibfnamefont {J.}~\bibnamefont
  {Jena}}, \bibinfo {author} {\bibfnamefont {B.}~\bibnamefont {G{\"o}bel}},
  \bibinfo {author} {\bibfnamefont {T.}~\bibnamefont {Ma}}, \bibinfo {author}
  {\bibfnamefont {V.}~\bibnamefont {Kumar}}, \bibinfo {author} {\bibfnamefont
  {R.}~\bibnamefont {Saha}}, \bibinfo {author} {\bibfnamefont {I.}~\bibnamefont
  {Mertig}}, \bibinfo {author} {\bibfnamefont {C.}~\bibnamefont {Felser}},\
  and\ \bibinfo {author} {\bibfnamefont {S.~S.~P.}\ \bibnamefont {Parkin}},\
  }\bibfield  {title} {\bibinfo {title} {Elliptical bloch skyrmion chiral twins
  in an antiskyrmion system},\ }\href
  {https://doi.org/10.1038/s41467-020-14925-6} {\bibfield  {journal} {\bibinfo
  {journal} {Nat. Commun.}\ }\textbf {\bibinfo {volume} {11}},\ \bibinfo
  {pages} {1115} (\bibinfo {year} {2020})}\BibitemShut {NoStop}%
\bibitem [{\citenamefont {Kim}\ \emph {et~al.}(2020)\citenamefont {Kim},
  \citenamefont {Kim}, \citenamefont {Kim}, \citenamefont {Kang}, \citenamefont
  {Shin}, \citenamefont {Lee}, \citenamefont {Min},\ and\ \citenamefont
  {Park}}]{Kim2020}%
  \BibitemOpen
  \bibfield  {author} {\bibinfo {author} {\bibfnamefont {J.}~\bibnamefont
  {Kim}}, \bibinfo {author} {\bibfnamefont {K.-W.}\ \bibnamefont {Kim}},
  \bibinfo {author} {\bibfnamefont {B.}~\bibnamefont {Kim}}, \bibinfo {author}
  {\bibfnamefont {C.-J.}\ \bibnamefont {Kang}}, \bibinfo {author}
  {\bibfnamefont {D.}~\bibnamefont {Shin}}, \bibinfo {author} {\bibfnamefont
  {S.-H.}\ \bibnamefont {Lee}}, \bibinfo {author} {\bibfnamefont {B.-C.}\
  \bibnamefont {Min}},\ and\ \bibinfo {author} {\bibfnamefont {N.}~\bibnamefont
  {Park}},\ }\bibfield  {title} {\bibinfo {title} {Exploitable magnetic
  anisotropy of the two-dimensional magnet cri3},\ }\href
  {https://doi.org/10.1021/acs.nanolett.9b03815} {\bibfield  {journal}
  {\bibinfo  {journal} {Nano Lett.}\ }\textbf {\bibinfo {volume} {20}},\
  \bibinfo {pages} {929} (\bibinfo {year} {2020})}\BibitemShut {NoStop}%
\bibitem [{\citenamefont {Webster}\ and\ \citenamefont
  {Yan}(2018)}]{PhysRevB.98.144411}%
  \BibitemOpen
  \bibfield  {author} {\bibinfo {author} {\bibfnamefont {L.}~\bibnamefont
  {Webster}}\ and\ \bibinfo {author} {\bibfnamefont {J.-A.}\ \bibnamefont
  {Yan}},\ }\bibfield  {title} {\bibinfo {title} {Strain-tunable magnetic
  anisotropy in monolayer ${\mathrm{crcl}}_{3}$, ${\mathrm{crbr}}_{3}$, and
  ${\mathrm{cri}}_{3}$},\ }\href {https://doi.org/10.1103/PhysRevB.98.144411}
  {\bibfield  {journal} {\bibinfo  {journal} {Phys. Rev. B}\ }\textbf {\bibinfo
  {volume} {98}},\ \bibinfo {pages} {144411} (\bibinfo {year}
  {2018})}\BibitemShut {NoStop}%
\bibitem [{\citenamefont {Xu}\ \emph {et~al.}(2020)\citenamefont {Xu},
  \citenamefont {Xie}, \citenamefont {Lu},\ and\ \citenamefont
  {Zhao}}]{XU2020126754}%
  \BibitemOpen
  \bibfield  {author} {\bibinfo {author} {\bibfnamefont {Q.-F.}\ \bibnamefont
  {Xu}}, \bibinfo {author} {\bibfnamefont {W.-Q.}\ \bibnamefont {Xie}},
  \bibinfo {author} {\bibfnamefont {Z.-W.}\ \bibnamefont {Lu}},\ and\ \bibinfo
  {author} {\bibfnamefont {Y.-J.}\ \bibnamefont {Zhao}},\ }\bibfield  {title}
  {\bibinfo {title} {Theoretical study of enhanced ferromagnetism and tunable
  magnetic anisotropy of monolayer cri3 by surface adsorption},\ }\href
  {https://doi.org/https://doi.org/10.1016/j.physleta.2020.126754} {\bibfield
  {journal} {\bibinfo  {journal} {Phys. Lett. A}\ }\textbf {\bibinfo {volume}
  {384}},\ \bibinfo {pages} {126754} (\bibinfo {year} {2020})}\BibitemShut
  {NoStop}%
\bibitem [{\citenamefont {Tang}\ \emph {et~al.}(2020)\citenamefont {Tang},
  \citenamefont {Zhang},\ and\ \citenamefont {Du}}]{D0TC04049E}%
  \BibitemOpen
  \bibfield  {author} {\bibinfo {author} {\bibfnamefont {C.}~\bibnamefont
  {Tang}}, \bibinfo {author} {\bibfnamefont {L.}~\bibnamefont {Zhang}},\ and\
  \bibinfo {author} {\bibfnamefont {A.}~\bibnamefont {Du}},\ }\bibfield
  {title} {\bibinfo {title} {Tunable magnetic anisotropy in 2d magnets via
  molecular adsorption},\ }\href {https://doi.org/10.1039/D0TC04049E}
  {\bibfield  {journal} {\bibinfo  {journal} {J. Mater. Chem. C}\ }\textbf
  {\bibinfo {volume} {8}},\ \bibinfo {pages} {14948} (\bibinfo {year}
  {2020})}\BibitemShut {NoStop}%
\bibitem [{\citenamefont {Tang}\ \emph {et~al.}(2023)\citenamefont {Tang},
  \citenamefont {Huang}, \citenamefont {Qin}, \citenamefont {Zhai},
  \citenamefont {Ideue}, \citenamefont {Li}, \citenamefont {Meng},
  \citenamefont {Nie}, \citenamefont {Wu}, \citenamefont {Bi}, \citenamefont
  {Zhang}, \citenamefont {Zhou}, \citenamefont {Chen}, \citenamefont {Qiu},
  \citenamefont {Tang}, \citenamefont {Zhang}, \citenamefont {Wan},
  \citenamefont {Wang}, \citenamefont {Liu}, \citenamefont {Tian},
  \citenamefont {Iwasa},\ and\ \citenamefont {Yuan}}]{Tang2023}%
  \BibitemOpen
  \bibfield  {author} {\bibinfo {author} {\bibfnamefont {M.}~\bibnamefont
  {Tang}}, \bibinfo {author} {\bibfnamefont {J.}~\bibnamefont {Huang}},
  \bibinfo {author} {\bibfnamefont {F.}~\bibnamefont {Qin}}, \bibinfo {author}
  {\bibfnamefont {K.}~\bibnamefont {Zhai}}, \bibinfo {author} {\bibfnamefont
  {T.}~\bibnamefont {Ideue}}, \bibinfo {author} {\bibfnamefont
  {Z.}~\bibnamefont {Li}}, \bibinfo {author} {\bibfnamefont {F.}~\bibnamefont
  {Meng}}, \bibinfo {author} {\bibfnamefont {A.}~\bibnamefont {Nie}}, \bibinfo
  {author} {\bibfnamefont {L.}~\bibnamefont {Wu}}, \bibinfo {author}
  {\bibfnamefont {X.}~\bibnamefont {Bi}}, \bibinfo {author} {\bibfnamefont
  {C.}~\bibnamefont {Zhang}}, \bibinfo {author} {\bibfnamefont
  {L.}~\bibnamefont {Zhou}}, \bibinfo {author} {\bibfnamefont {P.}~\bibnamefont
  {Chen}}, \bibinfo {author} {\bibfnamefont {C.}~\bibnamefont {Qiu}}, \bibinfo
  {author} {\bibfnamefont {P.}~\bibnamefont {Tang}}, \bibinfo {author}
  {\bibfnamefont {H.}~\bibnamefont {Zhang}}, \bibinfo {author} {\bibfnamefont
  {X.}~\bibnamefont {Wan}}, \bibinfo {author} {\bibfnamefont {L.}~\bibnamefont
  {Wang}}, \bibinfo {author} {\bibfnamefont {Z.}~\bibnamefont {Liu}}, \bibinfo
  {author} {\bibfnamefont {Y.}~\bibnamefont {Tian}}, \bibinfo {author}
  {\bibfnamefont {Y.}~\bibnamefont {Iwasa}},\ and\ \bibinfo {author}
  {\bibfnamefont {H.}~\bibnamefont {Yuan}},\ }\bibfield  {title} {\bibinfo
  {title} {Continuous manipulation of magnetic anisotropy in a van der {W}aals
  ferromagnet via electrical gating},\ }\href
  {https://doi.org/10.1038/s41928-022-00882-z} {\bibfield  {journal} {\bibinfo
  {journal} {Nat. Electron.}\ }\textbf {\bibinfo {volume} {6}},\ \bibinfo
  {pages} {28} (\bibinfo {year} {2023})}\BibitemShut {NoStop}%
\bibitem [{\citenamefont {Kim}\ and\ \citenamefont
  {Tserkovnyak}(2017)}]{PhysRevLett.119.077204}%
  \BibitemOpen
  \bibfield  {author} {\bibinfo {author} {\bibfnamefont {S.~K.}\ \bibnamefont
  {Kim}}\ and\ \bibinfo {author} {\bibfnamefont {Y.}~\bibnamefont
  {Tserkovnyak}},\ }\bibfield  {title} {\bibinfo {title} {Chiral edge mode in
  the coupled dynamics of magnetic solitons in a honeycomb lattice},\ }\href
  {https://doi.org/10.1103/PhysRevLett.119.077204} {\bibfield  {journal}
  {\bibinfo  {journal} {Phys. Rev. Lett.}\ }\textbf {\bibinfo {volume} {119}},\
  \bibinfo {pages} {077204} (\bibinfo {year} {2017})}\BibitemShut {NoStop}%
\bibitem [{\citenamefont {Li}\ \emph {et~al.}(2018)\citenamefont {Li},
  \citenamefont {Wang}, \citenamefont {Cao},\ and\ \citenamefont
  {Yan}}]{PhysRevB.98.180407}%
  \BibitemOpen
  \bibfield  {author} {\bibinfo {author} {\bibfnamefont {Z.-X.}\ \bibnamefont
  {Li}}, \bibinfo {author} {\bibfnamefont {C.}~\bibnamefont {Wang}}, \bibinfo
  {author} {\bibfnamefont {Y.}~\bibnamefont {Cao}},\ and\ \bibinfo {author}
  {\bibfnamefont {P.}~\bibnamefont {Yan}},\ }\bibfield  {title} {\bibinfo
  {title} {Edge states in a two-dimensional honeycomb lattice of massive
  magnetic skyrmions},\ }\href {https://doi.org/10.1103/PhysRevB.98.180407}
  {\bibfield  {journal} {\bibinfo  {journal} {Phys. Rev. B}\ }\textbf {\bibinfo
  {volume} {98}},\ \bibinfo {pages} {180407} (\bibinfo {year}
  {2018})}\BibitemShut {NoStop}%
\end{thebibliography}%

\end{document}